\documentclass[11pt, a4paper]{article}

\def\Komabanumber#1{\hfill \begin{minipage}{4.2cm}\tt UT-Komaba #1
\end{minipage}}

\title{\vspace{20mm}\bf{Construction of a Gauge-Invariant Action \\[2mm] for Type II Superstring Field Theory}\vspace{25mm}}
\author{\Large{Hiroaki Matsunaga}\footnote{E-mail:\,{\tt matsunaga@hep1.c.u-tokyo.ac.jp}} \vspace{3mm}}
\date{\it Institute of Physics, University of Tokyo \\[1mm] Komaba, Meguro-ku, Tokyo 153-8902, Japan \vspace{10mm}}

\usepackage{wrapfig}
\usepackage[dvips]{graphicx}
\usepackage{amsmath,amscd}
\usepackage[hypertex]{hyperref}
\usepackage[all]{xy}
\usepackage{cite}

\begin{document}

\maketitle

{\vspace{-128mm} \Komabanumber{13-04}\vspace{128mm}}

\begin{abstract}
We construct a gauge-invariant action for covariant type II string field theory in the NS-NS sector. 
Our construction is based on the large Hilbert space description and Zwiebach's string products are used. 
%
First, we rewrite the action for bosonic string field theory into a new form where a state in the kernel of the generator of the gauge transformation appears explicitly. 
Then we use the same strategy and write down our type II action, where a projector onto the small Hilbert space plays an important role. 
We present lower-order terms up to quartic order and show that three-point amplitudes are reproduced correctly. 
%
%
\end{abstract}

\thispagestyle{empty}

\clearpage

\tableofcontents

\thispagestyle{empty}

\newpage

\section{Introduction}

\setcounter{page}{1}

String field theory is one possible approach to understanding nonperturbative aspects of string theory \cite{Kaku:1974,Green:1984fu,Witten:1985cc,Hata:1986,Thorn:1988hm,LeClair:1988,Saadi:1989tb,Kugo:1989,Sen:1990ff,Schubert:1991en,Zwiebach:1992ie,Schwarz:1992nx,Kugo:1992md,Hata:1993gf,Sen:1993,Gaberdiel:1997ia}. 
In the bosonic theory, there exist two known Lorentz covariant theories: open string field theory \cite{Witten:1985cc} and closed string field theory \cite{Zwiebach:1992ie}. 
Open string field theory is described by the Chern-Simons-like action \cite{Witten:1985cc}. 
By contrast, closed string field theory necessitates an infinite number of fundamental vertices to cover the moduli space of Riemann surfaces and the action becomes a nonpolynomial form \cite{Saadi:1989tb, Kugo:1989}. 
However, underlying gauge structures of both theories are essentially equivalent: there exists some $nilpotent$ homotopy algebra which is related to poroperties of each moduli space. 
%
%
%
%
In free theory, using each BRST operator, we can construct the action for each string field theory and verify its gauge invariance in the same way. 
Hence, the underlying gauge structure of each free theory is described by the BRST-complex. 
Adding interaction terms requires us to extend or modify this framework. 
It is known that in full interaction theory, $A_{\infty }$-algebras for open strings,\footnote{
In fact, if we change the way of interaction from \cite{Witten:1985cc}, we cannot write down the cubic action \cite{Hata:1986}. 
Then, although it is open string field theory, the action has higher order vertices and its gauge structure is governed by $A_{\infty }$-algebras, which are yielded from properties of disk moduli space \cite{Nakatsu:2001da, Kajiura:2001ng}. 
} 
which includes differential graded algebras as subalgebras, and $L_{\infty }$-algebras for closed strings, which are extentions of Lie algebras equipped with a derivation, appear respectively. 
Using the nilpotency of $A_{\infty }$/$L_{\infty }$-algebras, we can easily check the gauge invariance, the equation of motion, and so on irrespective of the apparent complexity of the action. 
In this sense, we have good understandings of bosonic theory. 

\vspace{1mm}

By contrast, in superstring field theory, these geometrical and/or algebraical understandings are very little known. 
The formulation of superstring field theory has been in process and we have not constructed even a complete action until now. 
In addition to it, the situation assumes new aspects. 
In the NSR formalism of superstrings \cite{Friedan:1985ge, Polchinski:1998rr}, there exist ghost and picture number anomalies, which are one of difficulties in the formulaton of field theory: these anomalies make it difficult to construct supersymmetric theory as a naive extension of bosonic theory. 
To solve this problem, we often use the local insertion of picture changing operators but this leads to another problem: the collision of these operators causes a divergence. 
We would like to obtain the complete formulation of field theory of superstrings. 
In the construction of the concrete action, we have used two descriptions: the small Hilbert space description, which is based on superghosts $(\beta ,\gamma )$, and the Large Hilbert space description, which is based on bosonized superghosts $(\xi ,\eta ,\phi )$. 
The difference is whether state space does not include the zero-mode of $\xi$ or does. 
In the small Hilbert space description \cite{Witten:1986qs, Arefeva:1989cp, Preitschopf:1989fc}, the action always needs local insertions of picture changing operators and these operators causes difficulties: divergences, nontrivial kernels, and so on. 
By contrast, in the large Hilbert space description, the action does not always need local insertions of picture changing operators. 
However, it is not easy to construct the R-sector action.\footnote{
There exist some proposals of making R-sector theory. See, for example, \cite{Berkovits:2001im, Michishita:2004, Kroyter:2009}. 
} 
So far, limited to the NS sector, we know two full actions of superstring fields without picture changing operators: the action of open strings \cite{Berkovits:1995ab} and that of heterotic strings \cite{Okawa:2004ii}. 
In this paper, using the large Hilbert space description, we propose a construction of a classical action for type II string field theory in the NS-NS sector. 

\vspace{1mm}

In bosonic theory, from a geometrical point of view, the gauge structure is governed by $L_{\infty }$-algebras. 
We expect that in the NS-NS sector, the type II action has such properties like the bosonic action. 
(Of course, it is not clear yet what controls the algebraic poroperty of type II theory including all sectors. 
Recently, there has been a progress \cite{Jurco:2013qra}.) 
Therefore, we propose the following action, whose gauge structure is governed by $L_{\infty }$-algebras:
\begin{align}
S=\int_{0}^{1}{dt } \, \langle \partial _{t }\Psi (t ) , Q_{\mathcal{G}(t)} \Psi (t) \rangle _{\eta } .
\end{align}
It is constructed from type II string fields $\Psi (t)$ and a nilpotent operator $Q_{\mathcal{G}}$, which consists of type II string fields, the BRST operator, and the extension of Zwiebach's closed string products. 
As the inner product, we use the BPZ inner product with $\eta $-currents insertion: $\langle A,B\rangle _{\eta }:= \langle \eta _{0}^{+}\eta _{0}^{-}A,B \rangle$. 
The variation of this action, discussed in detail later, is given by the following form: 
\begin{align}
\delta S = \int_{0}^{1} {dt} \, \frac{\partial }{\partial t} \langle  \delta \Psi (t) , Q_{\mathcal{G}(t)} \Psi (t) \rangle _{\eta } .
\end{align}
Thus it has a simple gauge invariance: $\delta \Psi = Q_{\mathcal{G}}\Lambda$, which is yielded from $Q_{\mathcal{G}}^{2}=0$, and all information about the gauge structure described by $L_{\infty }$-algebras are encoded into $Q_{\mathcal{G}}$'s nilpotency. 
In this paper, we call this operator $Q_{\mathcal{G}}$ the BRST operator around $\mathcal{G}(t)$, which is deeply related to a pure gauge solution in bosonic theory. 
We would like to mention that our construction is based on algebraic properties of closed string fields, so we do not touch the geometrical understanding of superstrings, such as the correspondence of the full action and the decomposition of the moduli space of super-Riemann surfaces as the case of bosonic theory.

\vspace{2mm}

This article is organized as follows. 
In section \ref{2}, first, we briefly review the related results of bosonic string field theory \cite{Zwiebach:1992ie}. 
Then we rewrite the action into more suggestive form, which is easy to understand the gauge structure and its supersymmetric extension. 
This is one of new results. 
In section \ref{3}, we review the construction of a pure gauge solution of bosonic theory \cite{Schubert:1991en}. 
We pick up heterotic theory as an example and show the construction of the heterotic action using a (formal) bosonic pure gauge solution \cite{Okawa:2004ii}. 
At the end of this section, we put comments on a pure gauge solution of heterotic theory and the procedure of constructing type II theory. 
This section is devoted to present an idea of the construction of our type II action and does not include material which is necessary for reading other sections. 
After that, in section \ref{4} and sction \ref{5}, we propose the concrete form of the type II full action and see its properties: 
the gauge invariance and the equation of motion and so on. 
In paricular, we would like to show the correspondence of cohomology \cite{Kato:1982im}, the lower order action, and its gauge invariance through the perturbative expansion. 
We also check that it reproduces correct three point amplitudes which are correspond to the result of the first quantization theory \cite{Friedan:1985ge, AlvarezGaume:1988bg, Polchinski:1998rr} and estimate higher order amplitudes. 
At the top of section 2 and section 3, we pick up some formulae, which are necessary to follow practical calculations. 
In appendix \ref{Appendix A}, we summarize homotopy algebras \cite{Stasheff:1963, Getzler:1990, Stasheff:1993ny, Kimura:1993ea, Kontsevich:1997vb, Nakatsu:2001da, Kajiura:2001ng, Kajiura:2005sn, Getzler:2007}, which are used in this paper. 
In appendix \ref{Appendix B}, we present simple calculus of free theory using our new form of the action. 
It will be helpful for understanding this article because the same calculus goes in interacting theory. 
In Appendix \ref{Appendix C}, we consider open string field theory from the point of view of our construction. 
In this paper, any type closed string fields $\mathcal{A}$ all are imposed on the level matching condition $L_{0}^{-}\mathcal{A}=0$ and the subsidiary condition $b_{0}^{-}\mathcal{A}=0$. 

\section{Basic Facts of Closed String Field Theory} \label{2}
In this section, we review some results of bosonic theory \cite{Zwiebach:1992ie} and discuss several properties which are related to our construction of type II theory. 
In particular, we rewrite Zwiebach's action into another equivalent action which is rather clear to see the gauge structure: the generator of the gauge transformation, the correspondence between free and full theories, and so on. 

\vspace{2mm}

{\parindent=0pt { 
\underline{Relations between the BPZ Inner Product and String Products:} }}
\begin{align}
\langle A , B \rangle =& \, (-1)^{(A+1)(B+1)} \langle B ,A \rangle , \label{symmetry of BPZ}
\\
\langle Q A, B \rangle  =& \, (-1)^{A} \langle A , Q B \rangle , \label{Q and BPZ}
\\
\langle [A,B] ,C \rangle  =& \, (-1) ^{(A+B)} \langle A ,[ B,C ] \rangle . \label{cyclicity and BPZ}
\end{align}
\underline{$L_{\infty }$-identities:} ($l\geq 1$, $k\geq 2$, and $\sigma $ is the sign of splittings.) 
\begin{align}
\sum_{l+k=n} \sigma (i_{l},i_{k}) \big[ A_{i_{1}}, \dots ,A_{i_{l}} , [ A_{j_{1}} , \dots , A_{j_{k}} ] \big] =0 . \label{identities}
\end{align}
\underline{Stokes's Theorem of Three Point Vertices:} 
\begin{align}
\langle Q A , [ B, C ] \rangle 
+  (-1)^{A} \langle  A , [ QB , C ] \rangle + (-1)^{A+B}\langle A, [ B , QC ] \rangle =0  .\label{3pt}
\end{align}

\subsection{Zwiebach's Bosonic Closed String Field Theory}


The action of bosonic closed string field theory is described by closed string fields $\Psi$, the BPZ inner product, string products, and the BRST operator of closed strings $Q$. 
In particular, there is an infinite set of string products, all of which are graded-commutative. 
For example, the lowest product satisfies $[A, B ] = (-1)^{AB} [B, A]$, where string states in the exponent represent their Grassmann property, 0 (mod 2) for Grassmann even states and 1 (mod 2) for Grassmann odd states. 
This means, in closed string field theory, we need an infinite set of fundamental vertices \cite{Saadi:1989tb, Kugo:1989}.
The BPZ inner product is defined by BPZ conjugation as $\langle A,B \rangle := \langle A|c^{-}_{0} |B \rangle $, where $\langle A |$ is the BPZ conjugate of $|A\rangle $, and $c^{-}_{0}=\frac{1}{2}(c_{0}-\bar{c}_{0})$. 
It is nondegenerate on closed string Hilbert space. 
Using these components, the action of bosonic closed string field theory is given by 
\begin{align}
S_{B}:=\frac{1}{2} \langle \Psi , Q \Psi \rangle +\sum_{n\geq 1}\frac{\kappa ^{n}}{(n+2)!} \langle \Psi , [ \Psi ^{n} , \Psi ] \rangle  ,
\end{align}
where $\Psi$ is a bosonic closed string field which carries ghost number $2$, and $[ \Psi ^{n-1} ,\Psi ]$ is a closed string $n$-product\footnote{
We use a compact notation: $[\Psi ^{n}]:=[ \Psi , \dots , \Psi ]$. 
In paticular, $[\Psi ^{0}]:=0$ and $[\Psi ^{0} , \Psi ] :=[\Psi ] = Q\Psi $. 
} 
which carries ghost number $3+\sum_{i=1}^{n}(\mathrm{gh} (\Psi_{i})-2)$ \cite{Zwiebach:1992ie, Sen:1993}. 
Here, the symbol $\mathrm{gh}(\Psi _{i})$ means the ghost number of $\Psi _{i}$. 
The variation of this action is given by 
\begin{align}
\delta S_{B} =\sum_{n=0}^{\infty} \frac{\kappa ^{n}}{(n+1)!}  \langle \delta \Psi , [  \Psi ^{n} , \Psi ]   \rangle  .
\end{align}
Therefore, we obtain the following equation of motion: 
\begin{align}
\mathcal{F}(\Psi ) \equiv 
Q\Psi + \sum_{n\geq 1}\frac{\kappa ^{n}}{(n+1)!}[\Psi ^{n},\Psi ] =0 .
\end{align}
This action is invarint under the following gauge transformation: 
\begin{align}
\delta \Psi = \sum_{n=0}^{\infty} \frac{\kappa ^{n}}{n!} [ \Psi ^{n} , \Lambda ] .
\end{align}
The reason is, we can take resummention and use the following simple $L_{\infty }$-identities: 
\begin{align}
\sum_{l+k=n}\frac{n!}{l!k!}[ \Psi ^{l} , [ \Psi ^{k} ] ] =0     \hspace{10mm} ( l,k \geq 0 )  \,\, .
\end{align}
Note that the inner product has ciclicity and closed string products are symmetric. 
So we obtain $\delta S_{B} =0$ under such $\delta \Psi $, which is the gauge symmetry of closed string field theory.

\subsection{A Shifting Structure and $L_{\infty}$-algebras}
Let us consider the shift: $\Psi \rightarrow \Psi _{0}+\Psi ^{\prime }$.
If $S_{B}(\Psi )$ is invarinat under $\Psi \rightarrow \Psi +\delta \Lambda$, then clealy $S_{B}(\Psi _{0}+\Psi ^{\prime })$ is invariant under $\Psi _{0}+\Psi ^{\prime }\rightarrow \Psi _{0}+ \Psi ^{\prime } + \delta (\Psi _{0} +\Psi ^{\prime })$, and as a consequence it is invariant under $\Psi ^{\prime }\rightarrow \delta (\Psi _{0} +\Psi ^{\prime })$. This is the gauge invariance of the shifted action 
\cite{Sen:1990ff, Zwiebach:1992ie}. 
\begin{align}
\nonumber 
S_{B}( \Psi _{0}+ \Psi ^{\prime } ) &=\sum_{n=0}^{\infty }\frac{\kappa ^{n-2}}{n!}\sum_{m=0}^{n} 
\frac{n!}{m!(n-m)!} \{ \Psi ^{\prime m},\Psi _{0}^{n-m}  \} 
\\
&= \frac{1}{\kappa ^{2}}\sum_{n=0}^{\infty }\frac{\kappa ^{n}}{n!} \sum_{m=0}^{\infty }\frac{\kappa ^{m}}{m!} \{ \Psi ^{\prime n},\Psi_{0}^{m} \} 
\end{align}
Here, we use multilinear functions: $\{ B_{0}, B_{1}, \dots , B_{n} \} \equiv  \langle B_{0} , [ B_{1} , \dots , B_{n} ] \rangle $. 
We can define a new set of string products, denoted by the lowwer index of $\Psi _{0}$
\begin{align}
[ B_{1} , \dots , B_{n} ] _{\Psi _{0}} 
= \sum_{m=0}^{\infty } \frac{\kappa ^{m}}{m!} [ B_{1}, \dots , B_{n} , \Psi_{0}^{m}  ] ,
\end{align}
which are related to new multilinear functions as 
$\{ B_{0}, B_{1}, \dots , B_{n} \} _{\Psi _{0}} \equiv  \langle B_{0} , [ B_{1} , \dots , B_{n} ]_{\Psi _{0}} \rangle $. 

Using the definition of the lowwer-indexed ones, the shifted action simply reads
\begin{align}
S_{B}( \Psi _{0} +\Psi ^{\prime }) \equiv  S^{\prime }_{B}(\Psi ^{\prime } ) = \sum_{n=0}^{\infty }\frac{\kappa ^{n-2}}{n!} \{ \Psi ^{\prime n} \} _{\Psi _{0}} .
\end{align}
Here, the first two terms are not equal to zero for general $\Psi _{0}$: 
\begin{align}
\{ \Psi ^{\prime 0} \} = \sum_{m=2}^{\infty }\frac{\kappa ^{m}}{m!} \{ \Psi _{0}^{m} \} \equiv \kappa ^{2}S_{B}(\Psi _{0}) \, ,
\hspace{4mm}
\{ \Psi ^{\prime 1} \} = \sum_{m=1}^{\infty }\frac{\kappa ^{m}}{m!} \{ \Psi ^{\prime } , \Psi _{0}^{m} \} = \langle \Psi ^{\prime } , \kappa \mathcal{F}(\Psi _{0}) \rangle ,
\end{align}
where $\mathcal{F}(\Psi _{0})$ is equation of motion. Thus the shifted action, reads
\begin{align}
S_{B}(\Psi _{0}+ \Psi ^{\prime }) \equiv S^{\prime }_{B}(\Psi ^{\prime } ) =S(\Psi _{0} ) + \langle \Psi ^{\prime } , \mathcal{F}(\Psi _{0}) \rangle + \dots .
\end{align}
If and only if $\Psi _{0}$ satisfied the equation of motion: $\mathcal{F}(\Psi _{0})=0$, the term linear $\Psi ^{\prime }$ would have vanished. \cite{Zwiebach:1992ie}. 
The shifted action is invariant under the transformation: 
\begin{align}
\delta ^{\prime }\Psi ^{\prime } = \delta (\Psi _{0} +\Psi ^{\prime }) = \sum_{n=0}^{\infty }\frac{\kappa ^{n}}{n!} [ (\Psi _{0}+\Psi ^{\prime })^{n},\Lambda ] 
=\sum_{n=0}^{\infty } \frac{\kappa ^{n}}{n!} [ \Psi ^{\prime n} , \Lambda ]_{\Psi _{0}} ,
\end{align}
which is just the same form of the ordinary one. 
The first term 
is 
the new BRST-like operator, we call it the redefined BRST operator around $\Psi _{0}$: 
\begin{align}
[\Lambda ]_{\Psi _{0}} \equiv  Q_{\Psi _{0}}\Lambda := \sum_{n=0}^{\infty }\frac{\kappa ^{n}}{n!} [ \Lambda , \Psi _{0}^{n} ]=Q\Lambda +\sum _{n=1}^{\infty }[\Lambda , \Psi _{0} ^{n}] .
\end{align}
It does not have nilpotency for general $\Psi _{0}$ except for the case of classical solution: $\mathcal{F}(\Psi _{0})=0$.

\subsection{A New Action and the Gauge Invariance}
In the previous subsection, we reviewed the shifting structure of string field theory and considered new string products around constant string fields. 
However, algebraically, we are able to define new string products around an arbitrary string field $\phi$ as 
\begin{align}
\big[  A_{1}, \dots ,A_{m}  \big] _{\phi } := \sum_{n} \frac{\kappa ^{n}}{n!} \big[ \, \phi ^{n} ,A_{1}, \dots , A_{m} \big] .
\end{align}
There is a $L_{\infty }$-morphism between original products $[A_{1},\dots ,A_{n}]$ and new ones $[A_{1}, \dots , A_{m}]_{\phi }$. 
These new operators $Q_{\phi }\equiv [\, \cdot \, ]_{\phi }$ and $[A_{1},\dots ,A_{m}]_{\phi }$ have almost the same properties as those of old ones $Q$ and $[A_{1},\dots , A_{n}]$. 
(See \cite{Zwiebach:1992ie, Getzler:2007}, etc.) 
For instance, the nilpotency of $Q$ becomes 
\begin{align}
Q_{\phi } (Q_{\phi } \Psi ) = -\kappa [ \mathcal{F}(\phi ) , \Psi \, ]_{\phi } , \label{nilpotency on shell}
\end{align}
where $\mathcal{F}(\phi )$ is the equation of motion of bosonic closed string field theory. 
The most important fact is that these form a new $L_{\infty}$-algebra when $\phi $ satisfies the equation of motion.

\vspace{1mm}

Using these operators, we can rewrite the gauge transformation of bosonic string field theory $\delta \Psi = \sum{\frac{\kappa ^{n}}{n!}[\Psi ^{n} , \Lambda ]}$ into the simple form $\delta \Psi = Q_{\Psi }\Lambda $. 
As well as we identify the BRST operator $Q$ with the generator of the gauge transformation in free theory, we can regard this $Q_{\Psi}$ as the generator of the gauge transformation in full theory. 
We introduce the following notation: 
\begin{align}
 Q^{[a]}_{\Psi } := Q +\sum_{n}\frac{(a\cdot \kappa )^{n}}{n!} \big[ \, \Psi ^{n}, \hspace{3mm} \, \big] .
\end{align}
The upper index $[a]$ on $Q_{\Psi }$ means we consider closed string products of the coupling constant $a$ times $\kappa $. 
Then, $Q_{\Psi }^{[a]}$ gives a deformation of the gauge structure: $Q_{\Psi }^{[a]}$ connects $Q_{\Psi }^{[0]}=Q$ with $Q_{\Psi }^{[1]}=Q_{\Psi }$, where $a\in [0,1]$. 
Integrating this $Q_{\Psi }^{[a]}$ from $a=0$ to $a=1$, we can define a Maurer-Cartan operator $Q^{\prime }_{\Psi }$ as follows, 
\begin{align}
Q^{\prime }_{\Psi } := \int_{0}^{1}{da} \, Q^{[a]}_{\Psi } = \sum_{n=0}^{\infty } \frac{\kappa ^{n}}{(n+1)!} \big[ \, \Psi ^{n} , \hspace{3mm} \big] , \label{Q'}
\end{align}
which is deeply related to the action for string field theory, disucussed in the rest of this subsection. 
Note that the operator $Q_{\Psi }^{\prime }$ maps a string field $\Psi$ to a state $\mathcal{F}(\Psi )$ and the equation of motion is given by $\mathcal{F}(\Psi )= Q_{\Psi }^{\prime }\Psi = 0$. 

\vspace{1mm}

{\parindent=0pt{ \underline{Properties of the State $Q_{\Psi }^{\prime }\Psi $} }}

Before rewriting the action into a new form, we would like to see two useful properties, which are necessary to our calculus. 
The state $Q_{\Psi}^{\prime }\Psi $ has the following properties:
\begin{align}
Q_{\Psi }(Q_{\Psi }^{\prime } \Psi ) &=0 , \label{formula of QF}
\\ 
(-1)^{X}X (Q_{\Psi }^{\prime } \Psi ) &= Q_{\Psi } (X\Psi ) , \label{formula of XF}
\end{align}
where $X$ is a derivation which satisfies the relation\footnote{
In this paper, we always use the single bracket $[\Psi ^{n}]$ for string products, 
we therefore use the double bracket for the graded commutator: $[[A,B]]\equiv AB-(-1)^{AB}BA$. 
} $[[X , Q ]] =0$. 
The first line implies that the state $Q_{\Psi }^{\prime }\Psi $ belongs to the kernel of the generator $Q_{\Psi }$. 
It is a result from $L_{\infty }$-identities for an arbitrary string field $\psi$ whose ghost number is two:
\begin{align}
Q_{\psi }Q_{\psi }^{\prime }\psi 
=\sum_{k+l=n} \frac{\kappa ^{n}}{k!\cdot l!} \big[ \psi ^{k} , [ \psi ^{l} ] \big] =0 .
\end{align}
The second line means the $X$-derivative state $X(Q^{\prime }_{\Psi }\Psi)$ becomes the $Q_{\Psi }$-exact state. 
It is a result from the derivation propertiy of $X$ for string products $(-1)^{X}X[A^{n}]=n[A^{n-1}, XA ]$: 
\begin{align}
X ( Q_{\Psi }^{\prime }\Psi ) &= X \sum_{n=0}^{\infty } \frac{\kappa ^{n}}{(n+1)!} \big[ \Psi ^{n+1}  \big] 
\\
&= (-1)^{X} \sum_{n=0}^{\infty } \frac{\kappa ^{n}}{n!} \big[ \Psi ^{n} , X\Psi  \big] 
= (-1)^{X} Q_{\Psi }  (X\Psi ) .
\end{align}

\vspace{1mm}

{\parindent=0pt{ \underline{A New Action and the Gauge Invariance} }}

\vspace{1mm}

Using these operators, we can rewrite the action $S_{B}=\sum{\frac{\kappa ^{n}}{(n+2)!}}\langle \Psi , [\Psi ^{n} , \Psi ]  \rangle$ into a suggestive form. 
First, introducing real parameters $t\in [0,1]$ and $a\in [0,1]$, we obtain 
\begin{align}
\nonumber
S_{B}
&=  \int_{0}^{1}{dt} \,  \langle \partial_{t}(t \Psi )  ,\sum_{n=0}^{\infty} \frac{\kappa ^{n}}{(n+1)!} \big[ (t\Psi )^{n} , (t\Psi ) \big] \rangle 
\\
&= \int_{0}^{1}{dt} \int_{0}^{1}{da}\, \langle \partial_{t} (t \Psi ) ,  \sum_{n=0}^{\infty}  \frac{(a\kappa )^{n}}{n!} [ (t \Psi )^{n} ,(t\Psi ) ] \rangle .
\end{align}
Then, using (\ref{Q'}), we obtain the following action: 
\begin{align}
S_{B}&= \int_{0}^{1}{dt}  \int_{0}^{1}{da} \, \langle \partial_{t}(t \Psi ) , Q^{[a]}_{t\Psi } (t\Psi ) \rangle
= \int_{0}^{1}{dt} \, \langle \partial _{t}(t\Psi ) , {Q}_{t\Psi }^{\prime } (t\Psi ) \rangle . 
\end{align}
Or more formally, using $\Psi (t) $ which satisfies $\Psi (0)=0$ and $\Psi (1) = \Psi $, the action $S_{B}$ becomes  
\begin{align}
S_{B} =\int_{0}^{1}{dt} 
\langle  \partial _{t} \Psi (t) , Q_{\Psi (t)}^{\prime } \Psi (t) \rangle  .
\end{align}
This action has of course the same properties as the original action: the equation of motion $Q_{\Psi }^{\prime }\Psi =0$, the gauge invariance $\delta \Psi =Q_{\Psi }\Lambda $, and so on. 
Let us check these properties.

Using (\ref{Q and BPZ}), (\ref{symmetry of BPZ}), and (\ref{formula of XF}) for $X=\delta$ and $X= \partial _{t}$, the following relation holds: 
\begin{align}
\langle \partial _{t} \Psi , \delta \big( Q^{\prime }_{\Psi } \Psi \big) \rangle 
=\langle \partial _{t} \Psi , Q_{\Psi } \delta \Psi \rangle 
= \langle Q_{\Psi }\partial _{t} \Psi , \delta \Psi  \rangle 
=\langle \delta \Psi , Q_{\Psi } \partial _{t}\Psi \rangle 
= \langle \delta \Psi , \partial _{t} \big( Q^{\prime }_{\Psi }\Psi \big) \rangle .
\end{align}
Thus we can quickly calculate the variation $\delta S $ as follows: 
\begin{align}
\nonumber 
\delta S_{B}&=\int_{0}^{1}{dt} \, \Big(  \langle  \delta \big( \partial _{t}  \Psi (t) \big) , Q^{\prime }_{\Psi (t)} \Psi (t) \rangle + \langle \partial _{t} \Psi (t) , \delta \big( Q^{\prime }_{\Psi (t)} \Psi (t) \big) \rangle \Big) 
\\ \nonumber
&=\int_{0}^{1}{dt} \, \Big( \langle \partial _{t} \big( \delta \Psi (t) \big) , Q^{\prime }_{\Psi (t)} \Psi (t) \rangle 
+ \langle \delta \Psi (t) , \partial _{t} \big( Q^{\prime }_{\Psi (t)} \Psi (t) \big) \rangle \Big)  
\\
&= \int_{0}^{1}{dt}  \, \partial _{t} \, \langle \delta \Psi (t) , Q^{\prime }_{\Psi (t) }\Psi (t) \rangle 
= \langle \delta \Psi , Q_{\Psi }^{\prime } \Psi \rangle .
\end{align}
Then, using $L_{\infty }$-identities (\ref{formula of QF}), we obtain the expected gauge invariance $\delta \Psi = Q_{\Psi }\Lambda $ and the equation of motion $Q^{\prime }_{\Psi }\Psi =0$. 
We would like to emphasize that the variation of this type action is realized by exchanging the place of $\delta $ and $\partial _{t}$: 
\begin{align}
\delta S_{B}= \int_{0}^{1}{dt} \, \delta  \langle \partial _{t} \Psi (t) , Q^{\prime }_{\Psi (t)}\Psi (t) \rangle  
= \int_{0}^{1}{dt} \, \partial _{t} \, \langle \delta \Psi (t) , Q^{\prime }_{\Psi (t)}\Psi (t) \rangle . \label{delta S}
\end{align}
The statement that the variation is realized by exchanging the place of $\delta $ and $\partial _{t}$ is not limeted to the bosonic theory and it is correct in our type II theory. 
The result of (\ref{delta S}) also implies that using this representation, we are able to treat the full action $S_{B}=\int{dt}\langle \partial \Psi , Q_{\Psi }^{\prime }\Psi \rangle $ as the free action $S_{0}=\int{dt} \langle \partial \Psi , Q^{\prime } \Psi \rangle$ algebraically. 
(Cf. Appendix \ref{Appendix B})

\subsection{Another Calculation and Extra/BRST-exact Terms}
In the previous subsction, we obtain $\delta S_{B}$ from the relation $ \langle \partial _{t} \Psi , \delta \big( Q^{\prime }_{\Psi } \Psi \big)  \rangle = \langle \delta \Psi , \partial _{t} \big(  Q^{\prime }_{\Psi } \Psi  \big)  \rangle $ which is a result from (\ref{formula of XF}). 
To compare with the calculation of type II theory, it is helpfull to give a second look at this calculation: not using the property of the state $Q^{\prime }_{\Psi}\Psi$ but of the BRST-like operator $Q_{\Psi }^{[a]}$ for a derivation $X$ satisfying $[[X,Q]]=0$, namely,
\begin{align}
(-)^{X} X ( Q_{\Psi }^{[a]} \Psi ) = Q_{\Psi }^{[a]} (X\Psi )+ a\kappa \big[ X \Psi   , \Psi \big] ^{[a]}_{\Psi } . \label{formula2 of XF}
\end{align}
This relation is an $a$-integrand of the relation (\ref{formula of XF}). 
The upper index $[a]$ on $[A,B]_{\Psi }$ means we consider string products of the coupling constant $a$ times $\kappa $. 
The relation (\ref{formula2 of XF}) implies that for a derivation $X$ satisfying $[[X,Q]]=0$, the $X$-derivative state $X(Q_{\Psi }\Psi )$ becomes the BRST-exact state $Q_{\Psi }(X\Psi )$ plus extra terms $\kappa [ X\Psi , \Psi ]_{\Psi }$. 
Using (\ref{formula2 of XF}) for $X=\delta$ and $X=\partial _{t}$, we obtain the following calculation, which also goes in type II theory:
\begin{align}
\nonumber 
\delta S_{B} 
&=\int_{0}^{1}{dt}\int_{0}^{1}{da} \, \Big(  \langle \delta \partial _{t} \Psi  , Q^{[a]}_{\Psi  } \Psi  \rangle + \langle \partial _{t} \Psi  , \delta ( Q^{[a]}_{\Psi } \Psi ) \rangle \Big) 
\\ \nonumber 
&=\int_{0}^{1}{dt}\int_{0}^{1}{da} \, \Big( \langle \partial _{t} \delta \Psi  , Q^{[a]}_{\Psi } \Psi  \rangle 
+ \langle \partial _{t} \Psi ,  Q^{[a]}_{\Psi } \delta \Psi  \rangle
+ \langle \partial _{t}\Psi  , a\kappa [ \delta \Psi  , \Psi   ]^{[a]}_{\Psi } \rangle \Big) 
\\ \nonumber
&=\int_{0}^{1}{dt}\int_{0}^{1}{da} \, \Big( \langle \partial _{t} \delta \Psi  , Q^{[a]}_{\Psi } \Psi  \rangle 
+ \langle  \delta \Psi   , Q^{[a]}_{\Psi } \partial _{t} \Psi     \rangle
+  \langle \partial _{t}\Psi  , a\kappa [ \delta \Psi  , \Psi   ]^{[a]}_{\Psi } \rangle \Big) 
\\ \nonumber
&=\int_{0}^{1}{dt}\int_{0}^{1}{da} \, \Big( \langle \partial _{t} \delta \Psi  , Q^{[a]}_{\Psi } \Psi  \rangle 
+ \langle \delta \Psi  , \partial _{t} Q^{[a]}_{\Psi } \Psi  \rangle \Big) 
\\ \nonumber 
& \hspace{10mm} +  \int_{0}^{1}{dt}\int_{0}^{1}{da} \, \Big(  \langle \partial _{t}\Psi  , a\kappa [ \delta \Psi  , \Psi   ]^{[a]}_{\Psi } \rangle 
 -  \langle \delta \Psi  , a\kappa [ \partial _{t}\Psi  , \Psi   ]^{[a]}_{\Psi } \rangle \Big) 
\\
&=\int_{0}^{1}{dt} \int_{0}^{1}{da} \, \partial _{t} \, \langle \delta \Psi  , Q^{[a]}_{\Psi } \Psi  \rangle 
= \langle \delta \Psi  , Q^{\prime }_{\Psi } \Psi  \rangle . \label{bosonic calculation}
\end{align}
Although there appear extra terms of $\kappa [X\Psi , \Psi ]_{\Psi }$, in above calculation, using the cyclicity of the BPZ inner product, those of $X=\delta $ and $X=\partial _{t}$ cancel each other for general $\delta \Psi$.

We would like to mention that for $\delta \Psi = Q_{\Psi}\Lambda$, each extra term $\langle \partial _{t}\Psi , [\Psi ,Q_{\Psi}\Lambda ]_{\Psi }\rangle $ all vanishes itself on the mass shell because of $\int{da} \langle  Q_{\Psi}\Lambda , [ \Psi , \partial \Psi ]_{\Psi } \rangle \propto \langle \Lambda , [ \mathcal{F}(\Psi ) ,\partial \Psi ]_{\Psi } \rangle$. 
It is a result of (\ref{nilpotency on shell}) and/or essentially equivalent to the decoupling mechanism of BRST-exact terms, discussed in the rest of this subsection. 

\vspace{2mm}

{\parindent=0pt{\underline{Gauge Transformation as a Shift of an Expanding Point}}}

\vspace{1mm}

We can regard the gauge transformation $\delta \Psi = Q_{\Psi }\Lambda$ of the action $S_{B}$ as one kind of an infinitesimal shift of the expanding point of $Q_{\Psi }$, which is gauge equivalent to zero: 
\begin{align}
\int{dt} \int{da} \, \langle \partial _{t} \Psi , Q^{[a]}_{\Psi } \Psi \rangle \longrightarrow \int{dt} \int{da} \, \langle \partial _{t} \Psi , Q^{[a]}_{\Psi + \delta \Psi } \Psi \rangle .
\end{align}
The series of shifted action consists of various terms which have BRST-exact states. 
We would like to emphasize that those terms like $\langle A,[B,QC]\rangle $, which often appear in the series of shifted actions, are always vanish from the decoupling mechanism of BRST-exact terms. 
For example, we know $ \langle \partial _{t} \psi , [ \psi , Q \lambda ] \rangle  =0$ 
from $ \langle [ \partial _{t} \psi , \psi  ] , Q \lambda  \rangle 
= \langle [ Q \partial _{t} \psi , \psi ] + [ \partial _{t} \psi , Q \psi ] , \lambda \rangle $, (\ref{cyclicity and BPZ}), and (\ref{3pt}). 

When $\Psi$ satisfies the equation of motion $\mathcal{F}(\Psi )=0$, there exist an isomorphism of $L_{\infty}$-algebras between original one $(\mathcal{H},Q,[ \, \cdot \, ])$ and new one $(\mathcal{H},Q_{\Psi},[ \, \cdot \, ]_{\Psi })$. 
Thus, above extra terms in (\ref{bosonic calculation}) all vanish respectively on the mass shell.

\section{Pure Gauge Solutions and Supersymmetric Theory} \label{3}

In this section, we construct a pure gauge solution of bosonic string field theory and see its role in considering the supersymmetric extension of bosonic theory. 
As a simple example, we pick up heterotic string field theory \cite{Okawa:2004ii}. 
A pure gauge solution as a functional of superstring fields will also play an impotant role in the construction of type II theory. 
This section is devoted to present an idea and does not include material which is necessary for the following sections. 

\vspace{2mm}

{\parindent=0pt {
\underline{Action of $\eta _{0}$ for the BPZ Inner Product and String Products:} }}
\begin{align}
&\langle \eta _{0} A , \, B \rangle =(-1)^{A} \langle A , \, \eta _{0} B \rangle \, ,
\\
\eta _{0} [ A_{1} , \dots , A_{n} ] \, + &\sum_{i=1}^{n} (-1)^{(A_{1}+\dots +A_{i-1})} [ A_{1} , \dots , \eta _{0} A_{i} , \dots , A_{n} ] 
=0.
\end{align}

\subsection{A Classical Solution and Related Fields}



We can always construct a formal solution of bosonic closed string field theory $G(A)$ from closed string products and parameter fields $A$ which carries ghost number $1$ as follows:  
\begin{align}
G(A) := \int_{0}^{1}{d\tau } \, Q_{G(\tau A)} A .
\end{align}
Using $L_{\infty }$-identities, we can check quickly that the state $G(t)=G(A(t))$ satisfies the equation of motion $\mathcal{F}(G)=0$ in boconic closed string field theory: 
\begin{align}
\nonumber 
\mathcal{F}(G(A(t))) \equiv \,\, & QG(A(t)) +\sum_{n=1}^{\infty} \frac{k^{n}}{(n+1)!}[ G(A(t))^{n} , G(A(t)) ]
\\
= \, &\int_{0}^{1}{da} \left(  Q^{[a]}_{G(A(t))} G(A(t)) \right) =0 .
\end{align}
In this sence, we call $G(A(t))$ as a pure gauge solution\cite{Schubert:1991en, Kontsevich:1997vb, Kajiura:2001ng, Kajiura:2005sn,  Getzler:2007}.
It is equivalent to the result of \cite{Okawa:2004ii}: $G(A)$ is the solution of the following differential equation  
\begin{align}
\partial _{\tau }G(\tau A)=Q_{G}A:=QA+\sum_{n=1}^{\infty}\frac{\kappa^{n}}{n!}[G(\tau A)^{n},A]
\end{align}
with the initial condition $G(0)=0$. 
This equation means that such a pure gauge solution $G(A)$ can be built by successive infinitisimal gauge transformations path-dependently. 
The path is decided by choosing $\tau $-dependence of $A(t)$. 
As in the case of other gauge theories, we also choose a straight line connects $0$ and $A$, and parametrize the path linearly as $\tau A(t)$ with $0\leq \tau \leq 1$. 
After setting $\tau =1$, the first few terms of $G(A)$ are given by
\begin{align}
G(A)=QA+\frac{\kappa }{2} \big[ A,QA \big] +\frac{\kappa^{2}}{3!} \big[ A,QA,QA  \big] +\frac{\kappa^{2}}{3!} \big[ A, [A,QA]  \big]  +O(\kappa^{3}).
\end{align}
\underline{An Important Property of a Classical Solution}

\vspace{1mm}

The pure gauge solution has an important property: the state $G(A)$ is mapped into the $Q_{G}$-exact state by a derivation $X$ satisfying $[[X,Q]]=0$. 
In supersymmetric theory, we will use this property for practical calculations as $XQ^{\prime }_{\Psi}\Psi = (-1)^{X} Q_{\Psi }X\Psi$ in the previous section. 
Let us prove it quickly \cite{Okawa:2004ii}. 
This property is equivalent to the following statement: 
suppose that a pure gauge solution $G(A)$ and a derivation $X$ which satisfies $[[X,Q]]=0$ are given, then, we can always construct the state $H(A,X)$ which satisfies the following relation, 
\begin{align}
XG(A)=(-1)^{X}Q_{G}H(A,X) .
\end{align}
We call this state $H(A,X)$ ``a related field''. 
For instance, in bosonic closed string field theory, variation $\delta $ and derivation $\partial _{t}$ satisfy the above conditions of $X$, so there exist $H(A,\delta A)$ and $H(A,\partial _{t}A)$. 
%
%
We can obtain such $H(A,X)$ as the solution of the following differential equation:\footnote{
The reason becomes clear when we define the auxiliary field $\mathcal{H}(\tau )$ as
$\mathcal{H}(\tau ) := Q_{G(\tau A)} H(A,X ; \tau ) - (-1)^{X} X G(\tau A)  $. 
Then the result which we expect is equivalent to $\mathcal{H}(\tau =1)=0$. 
If our differencial equation is satisfied, we obtain $\partial _{\tau } \mathcal{H} = \kappa [ A, \mathcal{H} ]_{G}$. 
Using the initial condition $\mathcal{H}(\tau =0)=0$, it gives $\mathcal{H}=0$ for any $\tau$. 
} 
\begin{align}
\partial _{\tau }H(A,XA; \tau )=XA+\kappa [A, H(A,XA; \tau )]_{G} .
\end{align}
Then we can solve this in power of $\kappa $ as follows:
\begin{align}
H(A,XA)=XA+\frac{\kappa }{2} \big[ A,XA \big] +\frac{\kappa ^{2}}{3} \big[ A,QA,XA \big] +\frac{\kappa ^{2}}{3!} \big[ A,[A,XA] \big] +O(\kappa ^{3}) .
\end{align}

\subsection{Supersymmetric Theory and the Large Hilbert Space Description}
We reviewed the procedure of constructing a formal pure gauge solution $G(t)$ from parameter fields $A(t)$ which carries ghost number $(\mathrm{gh}[G]-1)$. 
In string field theory, pure gauge solutions of bosonic theories have played an impotant role in supersymmetric extension  \cite{Berkovits:1995ab, Okawa:2004ii}. 
At least it becomes algebraically clear when we consider in the large Hilbert space. 
In the large Hilbert space description, the ghost number (and of course the picture number) of superstring fields matches that of parameter fields of bosonic pure gauge solutions. 
We can construct an action of supersymmetric theory by identifying superstring fields with parameter fields of corresponding bosonic theory.

\vspace{1mm}

In the small Hilbert space description, the ghost number and the picture number of the string field $\Phi _{\mathrm{S}}$ are determinded by the natural correspondence to the vertex operator of strings of the first quantization theory. 
%
The difference between the large and small Hilbert space is whether we consider the zero mode of $\xi (z)$ or not. 
Thus we usually use the following identification \cite{Berkovits:1995ab} of the small space string field $\Phi _{S}$ and the large space string field $\Phi _{L}$: 
\begin{align}
\Phi _{\mathrm{S}} \cong  \eta _{0} \Phi _{\mathrm{L}}.
\end{align}
Here, $\eta _{0}$ is the zero mode of $\eta (z)$, which carries ghost number $1$ and picture number $-1$. 
In particular, it is a current which has conformal weight $1$ like the BRST operator. 
In the rest of this section, as an example of the supersymmetric extension, we pick up the construction of heterotic string field theory \cite{Okawa:2004ii} and see the role of the pure gauge solution. 
(cf. Appendix C.) 

\vspace{2mm}

{\parindent=0pt{
\underline{Heterotic String Field Theory}
}}

\vspace{1mm}

In heterotic string field theory \cite{Okawa:2004ii}, the small space string field $V_{\mathrm{S}}$ carries ghost number $2$ and picture number $-1$. 
Therefore, in the large space description, the heterotic string field $V$ carries ghost number $1$ and picture number $0$. 
It carries the same ghost number as above parameter fields $A(t)$, so we can built a pure gauge solution $G(t)$ by $V(t)$. 
Then the action is given by 
\begin{align}
S_{H}=\int_{0}^{1}{dt} \langle \eta_{0} \partial _{t} (tV) ,G(tV) \rangle ,
\end{align}
where we choose $t$-dependence linearly $V(t)=tV$. 
This is the Wess-Zumino-Witten-like action. 
Or more formally, this action can be written for general $t$-dependent $V(t)$ which satisfies $V(0)=0$ and $V(1)=V$, where $\Psi _{Q}:=G(V(t))$ and $\Psi _{X}:=H(V(t),X)$, as follows: 
\begin{align}
S_{H} =\int_{0}^{1} {dt }  \langle  \eta _{0}  \Psi _{t} , \Psi _{Q}  \rangle .
\end{align}
Note that for $X=(\partial _{t} , \delta  , \eta _{0})$, related fields $\Psi _{X}$ satisfy the following relations
\begin{align}
\partial _{t} \Psi _{Q} = Q_{\Psi _{Q}} \Psi _{t}
, \hspace{5mm} 
\delta \Psi _{Q} = Q_{\Psi _{Q}} \Psi _{\delta } 
, \hspace{5mm}  \eta _{0} \Psi _{Q} =-Q_{\Psi _{Q}}  \Psi _{\eta }  .
\end{align}
It is just the same result as that of open superstrings. 
The variation of this action is give by
\begin{align}
\delta S_{H} =\int_{0}^{1} {dt } \frac{\partial }{\partial t } \langle  \eta _{0} \Psi _{\delta } , \Psi _{Q}   \rangle .
\end{align}
So we obtain the equation of motion $\eta _{0} \Psi _{Q} =0 $ and the gauge invariance under 
\begin{align}
\Psi _{\delta } = Q_{\Psi _{Q}} \Lambda _{(0,0)} + \eta _{0} \Omega _{(0,1)},
\end{align}
where $\Lambda _{(0,0)}$ and $\Omega _{(0,1)}$ are gauge parameters with respect to $Q$ and $\eta _{0}$. 

We would like to mention that the $\xi _{0}$-decomposition of pure gauge solution $\Psi _{Q}$ is given by 
\begin{align}
\Psi _{Q} \equiv  \hat{\psi }_{Q} + \xi _{0} \hat{\psi }_{Q} = \hat{\psi }_{Q} + \xi _{0} ( -Q_{\Psi _{Q}} \Psi _{\eta } ) .
\end{align}
Using $\Psi _{\eta }:= H(V, \eta _{0}V) = \eta _{0}V+\dots $, we obtain the following form of the action for $V(t)=tV$: 
\begin{align}
S_{H}=\int_{0}^{1}{dt} \langle \eta _{0} \partial _{t}V, Q_{\Psi _{Q}} V \rangle + \dots .
\end{align}

\subsection{Pure Gauge Solutions of Heterotic String Field Theory}
To construct type II theory, it is not necessary to see contents of this subsection, but helpfull. 
As a pure gauge solution of bosonic theory $G(A)$ can always be constructed from parameter fields $A(t)$ which carry ghost number $\mathrm{gh}(G)-1$, we can also construct a pure gauge solution of heterotic theory $\mathcal{A}(\Phi )$ from parameter fields $\Phi (t)$ which carry ghost number $\mathrm{gh}(\mathcal{A})-1$.

\vspace{1mm}

{\parindent=0pt{
\underline{The BRST Operator Acting on Large Hilbert Space}
}}

\vspace{1mm}

We consider the action of the BRST operator $Q$ in the large Hilbert space. 
The $\xi$-zero mode decomposition of $\Psi$ is given by $\Psi=\hat{\psi } +\xi_{0}\hat{\phi } $. Here, $\hat{\psi}$ and $\hat{\phi}$ live in the small Hilbert space. 
\begin{align}
\nonumber 
Q\Psi =&Q(\hat{\psi} +\xi_{0}\hat{\phi } )
\\
=&\left( Q\hat{\psi }+\mathcal{X}_{0}\hat{\phi } \right)  +\xi_{0}\left( -Q\hat{\phi } \right) ,
\end{align}
where $\mathcal{X}_{0}:= [[ Q, \xi_{0} ]]$. 
Of course, the BRST operator $Q$ is nilpotent on the large Hilbert space:
\begin{align}
\nonumber 
Q^{2}\Psi =&Q\left(  Q\hat{\psi }+\mathcal{X}_{0}\hat{\phi } \right)  + Q \left( -\xi_{0}Q\hat{\phi }   \right) 
\\
  =& \left(  Q^{2}\hat{\psi} +Q\mathcal{X}_{0}\hat{\phi } \right) + \left( -\mathcal{X}_{0}Q\hat{\phi} +\xi_{0}Q^{2}\hat{\phi } \right) 
.
\end{align}
So we notice the suggestive fact about the nilpotency of the BRST operator: $Q(\mathcal{P} Q\Psi ) \not=0$, 
where $\mathcal{P}$ is the projector onto the small Hilbert space, which satisfies $\mathcal{P}+\mathcal{P}^{\perp }={\bf{1}}$ and $\mathcal{P}^{2}=\mathcal{P}$.
We know that the operator $Q$ is a linear operator on the small Hilbert spase $Q:\mathcal{H}_{S}\rightarrow \mathcal{H}_{S}$. 
So if the state $\hat{\psi }$ lives in the Kernel of $\eta _{0}$, the state $Q \hat{ \psi }$ also lives in the same space. 
By contrucs, the component $\xi _{0} \hat{ \phi }\in{\mathcal{H}_{L}}$ is mapped to elements of both components: $\mathcal{X}_{0}\hat{\phi}\in \mathcal{H}_{S}$ and $-\xi_{0}(Q\hat{\phi })\in \mathcal{H}_{L}$.

\vspace{2mm}

{\parindent=0pt{
\underline{Classical Solutions of Heterotic Theory and Type II String Fields}
}}

\vspace{1mm}

The equation of motion of heterotic theory is $\eta _{0} \Psi _{Q}=0$. 
It is helpfull for considering type II theory to construct a pure gauge solution of heterotic theory. 
Constructing a pure gauge solution $\mathcal{A}$ of heterotic theory is equivalent to finding $\Psi _{Q}$ which lives in the small Hilbert space. 
Using gauge parameter fields of heterotic theory $\Phi$, we can make such a field $\mathcal{A}$ quickly: $ \mathcal{A}(\Phi ):= \mathcal{P}^{\perp } \left( Q_{G} \Phi \right) 
$. 
The state $\mathcal{A}(t)$ lives in the large Hilbert space and $G( \mathcal{A} )$ belongs to the small Hilbert space because $\mathcal{P}^{\perp }=1-\mathcal{P}$ and $Q_{G}^{2}=0$. 
Thus it is a pure gauge solution. 

\vspace{1mm}

The state space of closed strings is composed from the tensor product of two sectors, those of right mover and left mover \cite{Friedan:1985ge, Polchinski:1998rr}. 
Heterotic string theory is the supersymmetric theory whose chiral sector is supersymmetric and the other is bosonic. 
There exist other supersymmetric theories that each sector has supersymmetry, so called type II theories. 
Through the analysis of this section, we know that a pure gauge solution plays an impotant role in supersymmetric extension and we notice the possibility that as we can always construct a formal pure gauge solution of bosonic string field theory as a functional of heterotic string fields \cite{Okawa:2004ii}, we would be able to construct it as a functional of type II string fields similarly: 
identification of gauge parameter fields of heterotic theory and type II string fields. 
In the next section, we demonstrate the construction of type II string field theory based on this perspective.

\section{Type II String Field Theory} \label{4}

Bosonic string field theory possesses the gauge symmetry generated by $Q_{\Psi_{B}}$. 
The generator $Q_{\Psi _{B}}$ has the kernel, which is an origin of the gauge symmetry: the state $Q^{\prime }_{\Psi _{B}} \Psi_{B}$. 
This state consists of bosonic string fields $\Psi _{B}$ and string products and 
satisfies $Q_{\Psi _{B}} Q^{\prime }_{\Psi _{B}} \Psi _{B} =0$ because of $L_{\infty }$-identities. 
Then a gauge invariant action 
 is constructed as follows
\begin{align}
S_{B}=\int_{0}^{1}{dt} \, \langle \partial _{t} \Psi _{B} (t) , Q_{\Psi _{B}}^{\prime }\Psi _{B}(t)\rangle .
\end{align}

We expect that the NS-NS-sector type II action also has such a gauge structure and assume that the generator is geven by $Q_{\mathcal{G}}$, where $\mathcal{G}=\mathcal{G}(\Psi )$ is a formal pure gauge solution of bosonic theory and carries ghost number $2$, which is a functional of type II string fields $\Psi$. 
The generator $Q_{\mathcal{G}}$ has the kernel: the state $Q_{\mathcal{G}(\Psi )}\Psi$. 
This state consists of type II string fields $\Psi$ and string products and satisfies $Q_{\mathcal{G}} Q_{\mathcal{G}(\Psi )}\Psi =0$ because of $L_{\infty }$-identities. 
Therefore, using this state, we propose the following classical action, which possesses the gauge symmetry generated by $Q_{\mathcal{G}}$: 
\begin{align}
S=\int_{0}^{1}{dt } \, \langle \partial _{t }\Psi (t ) , Q_{\mathcal{G}(t)} \Psi (t) \rangle _{\eta } .
\end{align}
It is constructed from string fields $\Psi (t)$, pure gauge solutions of bosonic theory $\mathcal{G}(t)=\mathcal{G}(\Psi (t))$  
and the BRST-like nilpotent operator $Q_{\mathcal{G}}$. 
As the inner product, we use the BPZ inner product with $\eta $-currents insertion: $\langle A,B\rangle _{\eta }:= \langle \eta _{0}^{+}\eta _{0}^{-}A,B \rangle$. 
They are discussed in subsection \ref{4.1} and \ref{4.2}. 

\vspace{1mm}

In this section, we give a concrete form of this action by constructing $\mathcal{G}=\mathcal{G}(\Psi )$ as a functional of type II string fields $\Psi$ and using properties of the state $Q_{\mathcal{G}}\Psi$ which are shown in subsection \ref{4.2}, the equation of motion and the gauge invariance of this action are presented.

\subsection{Type II String Fields and Pure Gauge Solutions}\label{4.1}

{\parindent=0pt{ \underline{Type II String Fields $\Psi$} }}

\vspace{1mm}

We write a type II string field as $\Psi$ and suppose it belongs to the large Hilbert space. 
Since we assume that $\eta_{0}^{+}\eta_{0}^{-}\Psi$ corresponds to the type II closed string vertex operator in superstring theory, the type II string field $\Psi $ carries ghost number $0$ and picture number $0$. 
Note that upper-indexed $\eta $-currents are given by $\eta_{0}^{\pm }:=\eta _{0}\pm \bar{\eta }_{0}$ and satisfy $\eta_{0}^{+}\eta _{0}^{-}\Psi =\eta_{0}\bar{\eta}_{0}\Psi $. 
Let us consider a path $\Psi (t )$ which has the starting point $0$ and the end point $\Psi$ on the state space of cloesd strings where $t \in [0,1]$ is a real parameter. 
Note that we can consider any $t$-dependent path as long as $\Psi (t )$ satisfies $\Psi (0)=0$ and $\Psi (1)=\Psi$. 
Using this $\Psi (t)$, we write down the type II action. 

\vspace{1mm}

In the rest of this subsection, we give one concrete construction of a pure geuge solution $\mathcal{G}$ and show an important property which general pure gauge solutions have. 

\vspace{2mm}

{\parindent=0pt
\underline{A Pure Gauge Solution $\mathcal{G}(\Psi )$ as a Functional of Type II String Fields $\Psi$}}

\vspace{1mm}

We would like to construct a pure gauge solution $\mathcal{G}=\mathcal{G}(\Psi )$ concretely, which is a functional of type II string fields $\Psi$. 
To this purpose, using Zwiebach's string products, we define $\mathcal{G}(\tau )$ as the solution of the following differential equation with the initial condition $\mathcal{G}(0)=0$: 
\begin{align}
\partial _{\tau }\mathcal{G} ( \tau ) = \sum_{n=0}^{\infty }\sum_{m=0}^{\infty } \frac{\kappa ^{n+m}}{n!m!} \big[  \mathcal{G}(\tau )^{n} , \mathcal{P}_{-} [ \mathcal{G}(\tau )^{m} , \Psi ]   \big] 
\equiv  Q_{\mathcal{G}}\mathcal{P}_{-}Q_{\mathcal{G}} \Psi ,
\end{align} 
where $\mathcal{P}_{-}$ is the projector\footnote{
Combining with $\mathcal{P}_{+}$, which is the projector onto $\mathrm{Im}[\eta _{0}^{+}]$, we can define the projector onto the closed string small Hilbert space as $\mathcal{P}_{+}\mathcal{P}_{-}$, namely, $\mathcal{P}_{- }\mathcal{P}_{+}:\mathcal{H}_{L}\otimes \mathcal{H}_{L}\rightarrow \mathcal{H}_{S}\otimes \mathcal{H}_{S}$. 
Of course, $\mathcal{P}_{+/-}$ satisfies 
$(\mathcal{P}_{+/-})^{2}=\mathcal{P}_{+/-}$. 
} onto $\mathrm{Im}[\eta _{0}^{-}]$, 
$\kappa $ is the coupling constant, and $\tau \in [0,1]$ is a real parameter. 
We can always obtain the solution $\mathcal{G}(\tau )$ by the formal Taylor series 
$\mathcal{G}(\tau )=\sum_{n=0}^{\infty } \frac{\tau ^{n}}{n!}\partial_{\tau }^{n}\mathcal{G}(0)$. 
This solution $\mathcal{G}(\tau )$ satisfies the equation of motion in bosonic theory for any $\tau$. 
It is a result from the uniqueness of the solution of the following first-order differential equation: 
\begin{align}
\partial _{\tau }\mathcal{F}(\mathcal{G}(\tau )) &= Q \partial _{\tau } \mathcal{G}(\tau ) + \sum_{n=1}^{\infty }\frac{\kappa ^{n}}{(n+1)!} \partial _{\tau } [ \mathcal{G}(\tau )^{n} , \mathcal{G}(\tau ) ]
\\
&=Q_{\mathcal{G}} \partial_{\tau } \mathcal{G}(\tau ) = Q_{\mathcal{G}}^{2} \left( \mathcal{P}_{-}Q_{\mathcal{G}}\Psi \right) = \kappa \big[  \mathcal{P}_{-}Q_{\mathcal{G}}\Psi , \mathcal{F}(\mathcal{G}(\tau )) \big] _{\mathcal{G}} .
\end{align}
Using the initial condition $\mathcal{F}(\mathcal{G}(0))=0$, we obtain $\mathcal{F}(\mathcal{G}(\tau ) )=0$ for any $\tau$, 
which is the solution of $\partial _{\tau }\mathcal{F} = \kappa [ \mathcal{P}_{-}Q_{\mathcal{G}}\Psi , \mathcal{F} ]$. 
Thus $\mathcal{G}(\tau )$ is a pure gauge solution, and we define $\mathcal{G}$ as $\mathcal{G}:=\mathcal{G}(1)$. 

The first few terms of $\mathcal{G}$ are calculated as follows: 
\begin{align}
\mathcal{G}(\Psi )&= Q\mathcal{P}_{-}Q\Psi 
+ \frac{\kappa }{2}\Big(  \big[ Q\mathcal{P}_{-}Q\Psi , \mathcal{P}_{-}Q\Psi \big] + Q\mathcal{P}_{-} \big[ Q\mathcal{P}_{-}Q\Psi , \Psi \big]  \Big)
\\ \nonumber
& \hspace{4mm} +\frac{\kappa ^{2}}{3!} \Big(  \big[ \big( Q\mathcal{P}_{-}Q\Psi \big) ^{2}, \mathcal{P}_{-} Q \Psi \big] + \big[  [ Q\mathcal{P}_{-}Q\Psi , \mathcal{P}_{-}Q\Psi ] +  Q\mathcal{P}_{-}[ Q\mathcal{P}_{-}Q\Psi  , \Psi  ]  , \mathcal{P}_{-}Q\Psi \big] 
\\ \nonumber
& \hspace{5mm} 
+ Q\mathcal{P}_{-} \Big(  \big[ \big( Q\mathcal{P}_{-}Q\Psi \big) ^{2}, \Psi \big] + \big[  [ Q\mathcal{P}_{-}Q\Psi , \mathcal{P}_{-}Q\Psi ] + Q\mathcal{P}_{-}[Q\mathcal{P}_{-}Q\Psi , \Psi ]  , \Psi \big]  \Big) \Big)
+\dots .
\end{align}
We succeeded to construct a pure gauge solution $\mathcal{G}$ as a functinal of type II string fields $\Psi$. 
Using it, we introduce a function $\mathcal{G}(t)\equiv \mathcal{G}(\Psi (t))$, which is of course a pure gauge solution for any $t$. 
Note that the $t$-dependence of $\mathcal{G}(t)$ is determineded by that of $\Psi (t)$ and we can take it arbitrary as long as $\Psi (t)$ satisfies $\Psi (0)=0$ and $\Psi (1)=\Psi $. 

\vspace{2mm}

{\parindent=0pt 
\underline{A Important Property of Pure Gauge Solutions and Variation $\delta \Psi = Q_{\mathcal{G}}\Lambda$} }

\vspace{1mm}

The pure gauge solution has an important property: the state $\mathcal{G}(\Psi )$ is mapped into the $Q_{\mathcal{G}}$-exact state by a derivation $X$ satisfying $[[X,Q]]=0$. 
In other words, suppose the state $\mathcal{G}(\Psi )$ satisfying $\mathcal{F}(\mathcal{G}(\Psi ))=0$ and some derivation $X$ satisfying $[[ X , Q ]] =0$ are given, then there exists a state $\Lambda _{X}(\Psi )$ which satisfies the following relation: 
(We also call this $\Lambda _{X}$ ``a related field''.)
\begin{align}
X \mathcal{G}(\Psi ) = (-1)^{X} Q_{\mathcal{G}} \Lambda _{X}(\Psi ) . \label{XG}
\end{align}
This $\Lambda _{X}(\tau )$ is also a functional of $\Psi$. 
In paticular, $\Lambda _{X}(0)=0$ and we define $\Lambda _{X}$ as $\Lambda _{X}:=\Lambda _{X}(1)$. 
As well as the case in the previous section, we can construct this $\Lambda _{X}(\tau )$ concreatly as the solution of the first-order differential equation $\partial _{\tau } \Lambda _{X}(\tau ) = X\mathcal{P}_{-}Q_{\mathcal{G}}\Psi +\kappa [ \mathcal{P}_{-}Q_{\mathcal{G}}\Psi , \Lambda _{X}(\tau ) ]_{\mathcal{G}}$ with the initial condition $\Lambda _{X}(0)=0$. 
The most interesting one is the related field $\Lambda_{\delta }$ associated with the variation $\delta \Psi = Q_{\mathcal{G} }\Lambda $. 
In the rest of this subsection, we see useful properties of $\Lambda _{X}(\tau )$, which is helpfull to our calculus. 
Suppose this derivation $X$ commutes with the projector $\mathcal{P}_{-}$, then we can calcurate $\partial _{\tau }\Lambda _{X}(\tau )$ as 
\begin{align}
\partial_{\tau } \Lambda _{X} 
=(-1)^{X} \mathcal{P}_{-} \Big(  Q_{\mathcal{G}} X\Psi + \kappa [X\mathcal{G} , \Psi ]_{\mathcal{G}}
\Big) +\kappa [ \mathcal{P}_{-}Q_{\mathcal{G}}\Psi , \Lambda_{X} ]_{\mathcal{G}} .
\end{align}
For $\delta \Psi = Q_{\mathcal{G}}\Lambda$, it becomes 
$\partial _{\tau } \Lambda _{\delta }(\tau ) = \kappa \big( \mathcal{P}_{-} [ Q_{\mathcal{G}}\Lambda_{\delta }(\tau ) , \Psi ]_{\mathcal{G}}+ [ \mathcal{P}_{-}Q_{\mathcal{G}}\Psi , \Lambda _{\delta } ]_{\mathcal{G}}\big) $. 
Thus we obtain $\delta \mathcal{G}(\tau )=Q_{\mathcal{G}}\Lambda_{\delta }(\tau )=0$ under $\delta \Psi =Q_{\mathcal{G}}\Lambda$ because $\Lambda _{\delta }(\tau )=0$ for any $\tau$ when $Q_{\mathcal{G}}\delta \Psi =0$.

\subsection{The Generator $Q_{\mathcal{G}}$ and the State $Q_{\mathcal{G}}\Psi $} \label{4.2}

In the previous section, we constructed $\mathcal{G}(t)=\mathcal{G}(\Psi (t))$ as a functional of type II string fields $\Psi$, which carries the same ghost number as a bosonic string field. 
Therefore we can define the BRST-like operator $Q_{\mathcal{G}(t)}$ as an operator which depends on type II string fields $\Psi (t)$ through the functional $\mathcal{G}=\mathcal{G}(\Psi )$, which will become the generator of the gauge transformation of our type II action. 
The generator $Q_{\mathcal{G}}$ has the kernel: the state
\begin{align}
Q_{\mathcal{G}(t)} \Psi (t) \equiv  Q\Psi (t) +\sum_{n=1}^{\infty} \frac{ \kappa ^{n}}{n!} \big[ \mathcal{G}(t)^{n} ,\Psi (t) \big] , 
\end{align}
which satisfies $Q_{\mathcal{G}}Q_{\mathcal{G}(\Psi )}\Psi =0$ at $t=0$. 
In particular, the state $Q_{\mathcal{G}}\Psi$ has the following properties:
\begin{align}
Q_{\mathcal{G}}(Q_{\mathcal{G}}\Psi )&=0 , \label{nilpotency}
\\
(-1)^{X}X(Q_{\mathcal{G}}\Psi )&=Q_{\mathcal{G}}(X\Psi ) + \kappa [ Q_{\mathcal{G}}\Lambda _{X} , \Psi ]_{\mathcal{G}} , \label{XQg}
\\
\langle \Psi , X(Q_{\mathcal{G}}\Psi )\rangle &=(-1)^{X}\langle \Phi , Q_{\mathcal{G}}(X\Psi )\rangle , \label{XQg in BPZ}
\end{align}
where $X$ is a derivation which satisfies $[[X,Q]]=0$ and $\Phi$ is an arbitraly string field. 
These properties are used in later calculations, so let us prove them. 
The first line means the nilpotency of the generator $Q_{\mathcal{G}}$. 
Since $\mathcal{G}(\Psi )$ is a pure guage solution, using (\ref{identities}), we can check it as follows: 
\begin{align}
Q_{\mathcal{G}}Q_{\mathcal{G}}\Psi = \sum_{n,m} \sum_{\sigma } \frac{\kappa ^{n+m}}{n!(m+1)!} \big[ \mathcal{G}^{n} , \sigma \big( [ \mathcal{G}^{m} , \Psi ] \big)  \big] =
-\kappa \big[ \Psi , \mathcal{F}(\mathcal{G}) \big] _{\mathcal{G}} =0 ,
\end{align}
where $\sigma ([A_{1},\dots , A_{n}])$ is a permutation of the order of $[A_{1}, \dots  ,A_{n}]$ and the sum of $\sigma $ runs over all possible permutations.  
%
%
The second line implies the state $Q_{\mathcal{G}}\Psi $ is mapped into the $Q_{\mathcal{G}}$-exact state plus extra terms by a derivation $X$ and the third line guarantees that these extra terms all vanish in the BPZ inner product. 
Using the property (\ref{XG}), we can prove the property (\ref{XQg}) as same as (\ref{formula2 of XF}). 
Note that extra terms all are the form $\langle \Phi , [ \Psi , Q_{\mathcal{G}} \Lambda _{X} ]_{\mathcal{G}} \rangle $ and it is sufficient for the proof of (\ref{XQg in BPZ}) to give $\langle \Phi , [ \Psi , Q_{\mathcal{G}}\Lambda ]_{\mathcal{G}} \rangle =0$. 
The proof becomes understandable by separating two cases: whether $X$ is even or odd. 
When $X$ is even, namely, $(-1)^{X}=1$, the Grassmann parity of $\Phi$ becomes even. 
Then, by (\ref{cyclicity and BPZ}), (\ref{Q and BPZ}), (\ref{identities}), and (\ref{symmetry of BPZ}), the following calculation goes :
\begin{align}
\langle \Phi &, [ \Psi , Q_{\mathcal{G}} \Lambda _{X} ]_{\mathcal{G}} \rangle 
= \langle [ \Phi , \Psi ]_{\mathcal{G}} , Q_{\mathcal{G}} \Lambda _{X} \rangle 
=\langle [ Q_{\mathcal{G}} \Phi , \Psi ]_{\mathcal{G}} , \Lambda _{X}\rangle  + \langle [ \Phi , Q_{\mathcal{G}} \Psi ]_{\mathcal{G}} , \Lambda _{X} \rangle 
\\ \nonumber 
&=\langle [ Q_{\mathcal{G}} \Phi , \Lambda _{X} ]_{\mathcal{G}} , \Psi \rangle  + \langle [ Q_{\mathcal{G}} \Psi , \Lambda _{X}  ]_{\mathcal{G}} , \Phi \rangle 
=\langle  Q_{\mathcal{G}} \Phi , [ \Lambda _{X} , \Psi ]_{\mathcal{G}} \rangle  + \langle \Phi , [ Q_{\mathcal{G}} \Psi , \Lambda _{X}  ]_{\mathcal{G}}  \rangle .
\end{align}
We therefore obtain $\langle \Phi , [ \Psi , Q_{\mathcal{G}} \Psi _{X} ]_{\mathcal{G}} \rangle =0$ because of (\ref{3pt}) and 
\begin{align}
\langle \Phi , [ \Psi , Q_{\mathcal{G}}\Lambda _{X} ]_{\mathcal{G}} \rangle 
= \frac{1}{2} \Big( \langle  Q_{\mathcal{G}} \Phi , [ \Psi , \Psi _{X} ]_{\mathcal{G}} \rangle 
+ \langle \Phi , [ Q_{\mathcal{G}} \Psi , \Lambda _{X} ]_{\mathcal{G}} \rangle 
+ \langle \Phi , [ \Psi , Q_{\mathcal{G}} \Lambda _{X} ]_{\mathcal{G}} \rangle  \Big) .
\end{align}
For odd $X$, the parity of $\Phi$ becomes odd and we can check (\ref{XQg in BPZ}) similarly. 
This is equivalent to the decoupling mechanism of BRST-exact states in bosonic string field theory.\footnote{
Or more simply, these extra terms of $ \delta \mathcal{G} = Q_{ \mathcal{G} } \Lambda _{ \delta }$ and $ \eta \mathcal{G} = - Q_{ \mathcal{G}} \Lambda _{\eta }$ are equivalent to the gauge-trivail shift of $Q_{\mathcal{G}}$'s expanding point: $Q_{\mathcal{G}}\mapsto Q_{\mathcal{G}+Q_{\mathcal{G}}\lambda}$. Thus we can expect that the result is not affected by these terms. 
}  

\vspace{1mm}

We would like to mention that in our construction, although the operator $Q_{\mathcal{G}}$ anticommute with $\eta _{0}^{-}$, it dose not anticommute with $\eta _{0}^{+}$ as $[[  Q_{\mathcal{G} }  , \eta _{0}^{+} ]]  =  \kappa [ \eta _{0}^{+} \mathcal{G} , \hspace{2mm}  ]_{\mathcal{G}}$. 
However, using $\eta _{0}^{+}\mathcal{G}=-Q_{\mathcal{G}}\Lambda _{\eta }$ and (\ref{XQg in BPZ}), this commutator becomes an element of the kernel of cyclic $L_{\infty }$-algebras. 
As a result, the operator $Q_{\mathcal{G}}$ and $\eta _{0}^{+}$ anticommute in the BPZ inner product. 
We therefore treat the classical solution $\mathcal{G}$ as it belongs to the small Hilbert space in the BPZ inner product. 
Thus, for example, we obtain
$\langle \eta _{0}^{-}\partial _{t} \Psi , \eta_{0}^{+} \big( Q_{\mathcal{G}}\Psi \big) \rangle 
=-\langle \eta_{0}^{-}\partial _{t} \Psi , Q_{\mathcal{G}} \big( \eta _{0}^{+}\Psi \big) \rangle$. 
This mechanism admits us to use a simple inner product in which $Q_{\mathcal{G}}$ works as a derivation: the BPZ inner product with $\eta$-currents insertion $\langle A,B\rangle _{\eta }:=\langle \eta _{0}^{+} \eta _{0}^{-} A , B \rangle$. 
It is nonzero if and only if the total ghost and picture number of the states in the inner product are equal to $1$ and $0$ respectively. 
\begin{align}
\langle A , Q_{\mathcal{G}} B \rangle _{\eta } = (-1)^{A} \langle Q_{\mathcal{G}} A, B\rangle _{\eta } . \label{Qg in BPZ}
\end{align}

\subsection{The Classical Action and the Gauge Invariance}

Using operators defined in the previous subsection, we propose the following NS-NS-sector action for type II string field theory in the large Hilbert space description: 
\begin{align}
S= \int_{0}^{1}{dt} \, \langle \partial _{t} \Psi (t) , Q_{\mathcal{G}(t)} \Psi (t) \rangle _{\eta } \equiv 
\int_{0}^{1}{dt} \,  \langle \eta^{+}_{0} \eta^{-}_{0} \partial _{t} \Psi (t) , Q_{\mathcal{G}(t)} \Psi (t) \rangle .
\end{align}
The $t$-dependence of $\mathcal{G}(t)=\mathcal{G}(\Psi (t))$ is determinded by that of $\Psi (t)$, and 
we can take $\Psi (t)$ as an arbitrary function of $t\in [0,1]$ as long as it satisfies $\Psi (0) =0$ and $\Psi (1)=\Psi $. 
In practical calculations, we often take $t$-dependence linearly $\Psi (t) =t\Psi $ for simplicity. 
This action is almost the same form as the bosonic action. 
In fact, using (\ref{XQg in BPZ}) for $X=\delta , \partial _{t}$ and (\ref{Qg in BPZ}), we can carry out calculations as same as bosonic theory. 
Note that the following calculation goes: 
\begin{align}
\langle \partial _{t} \Psi, \delta ( Q_{\mathcal{G}} \Psi ) \rangle _{\eta } 
= \langle \partial _{t} \Psi , Q_{\mathcal{G}} ( \delta \Psi ) \rangle _{\eta }
= \langle Q_{\mathcal{G}} ( \delta \Psi ) , \partial _{t} \Psi \rangle _{\eta }
= \langle \delta \Psi , Q_{\mathcal{G}} ( \partial _{t} \Psi ) \rangle _{\eta }
= \langle \delta \Psi , \partial _{t} ( Q_{\mathcal{G}} \Psi ) \rangle _{\eta } .
\end{align}
Hence, the variation of the action is given by:
\begin{align}
\nonumber 
\delta S &= \int_{0}^{1}{dt} \, \Big( \langle \partial _{t} ( \delta \Psi (t) ) , Q_{\mathcal{G}(t)} \Psi (t)  \rangle _{\eta } 
+ \langle \partial _{t}\Psi (t) , \delta (Q_{\mathcal{G}(t)} \Psi (t) ) \rangle _{\eta } \Big) 
\\ \nonumber 
&= \int_{0}^{1}{dt} \, \Big(  \langle \partial _{t} ( \delta \Psi (t) ) , Q_{\mathcal{G}(t)} \Psi (t)  \rangle _{\eta } 
+ \langle \delta \Psi (t) , \partial _{t} ( Q_{\mathcal{G}(t)} \Psi (t) ) \rangle _{\eta } \Big) 
\\
&= \int_{0}^{1}{dt} \,  \partial _{t} \, \langle \delta \Psi (t) , Q_{\mathcal{G}(t)} \Psi (t)  \rangle _{\eta } 
= \langle \delta \Psi , Q_{\mathcal{G}}\Psi  \rangle _{\eta }  .
\end{align}
As a result, we obtain the gauge invariance under the gauge transformation $\delta \Psi = Q_{\mathcal{G}}\Lambda $. 
As an element of the large Hilbert space, the gauge transformation can be writen down as follows: 
\begin{align}
\delta \Psi = Q_{\mathcal{G}} \Lambda + \eta ^{+}_{0} \Omega ^{+} +\eta ^{-}_{0} \Omega ^{-} .
\end{align}
Here, $\Lambda$, which carries ghost number $-1$ and picture number $0$, and $\Omega ^{\pm }$, which carries ghost number $-1$ and picture number $1$, are gauge parameters with respect to $Q_{\mathcal{G}}$ and $\eta _{0}^{\pm }$.

The equation of motion of our type II theory is given by
\begin{align}
Q_{\mathcal{G}} \eta ^{+}_{0} \eta ^{-}_{0}  \Psi =\eta _{0}^{+}\eta _{0}^{-}Q_{\mathcal{G}}\Psi =0 .
\end{align}
In particular, on the mass shell, we obtain $\eta _{0}^{\pm }\mathcal{G}(\Psi )=0$, which is the result of the equation of motion. 
We can obtain all interaction terms from this action. In the next section, we calculate the expansion with respect to $\kappa $ and see that it gives correct three point amplitudes.

\subsection{Summary} \label{4.4}
We proposed the following type II full action in the large Hilbert space description, 
\begin{align}
S=\int_{0}^{1}{dt} \, \langle \eta^{+}_{0}\eta^{-}_{0} \partial _{t} \Psi (t) , Q_{\mathcal{G}(t)}\Psi (t) \rangle  .
\end{align}
In this theory, $Q_{\mathcal{G}}$ and $\eta _{0}^{\pm }$ commute each other in the BPZ inner product and the nonlinear transformations with respect to $\eta _{0}^{\pm }\Omega ^{\pm }$ do not appear. 
Therefore, we can include $\eta ^{\pm }_{0}$ into the definition of the innner product as 
$\langle A, B\rangle _{\eta } := \langle \eta _{0}^{+} \eta _{0}^{-} A , B \rangle $. 
Of course, the inner product $\langle A , B \rangle _{\eta }$ is degenerate in the large Hilbert space $\mathcal{H}_{L}$: all of the small Hilbert space components are identified with null states. 
Thus we consider the equivalence relation $A\sim A+\eta_{0}^{+}\Omega ^{+} + \eta _{0}^{-} \Omega ^{-}$, and introduce the quotient space $\mathcal{H}_{\eta }:= \mathcal{H}_{L}/ \mathrm{Ker}[\eta _{0}^{\pm }]$. 
In this quotient space, the inner product $\langle A, B\rangle _{\eta }$ becomes nondegenerate. 
This situation is similar to the $c_{0}^{-}$-insertion (considering $b_{0}^{-}$-vanishing states) in bosonic closed string field theory. 
Using the quotient space $\mathcal{H}_{\eta }$ as the state space, we can write down the action as follows: 
\begin{align}
S=\int_{0}^{1}{dt} \, 
\langle \partial _{t} \Psi (t) , Q _{\mathcal{G}} 
\Psi (t) \rangle _{\eta } .
\end{align}
Here, we regard $\Psi (t)$ as an element of $\mathcal{H}_{\eta }$. 
Then this action is invariant under the gauge transformation $\delta \Psi =Q_{\mathcal{G}}\Lambda$. 
The equation of motion becomes $Q_{\mathcal{G}}\Psi =0$. 
Note that while the symmetric prorerties of string products also hold, the naive cyclicity of the inner product is lost in this discription. (It is natural because $\mathcal{G}(\Psi )$ is equivalent to $0$ in $\mathcal{H}_{\eta }$.)

This representation of the type II action will give us an one-to-one correspondence to the small Hilbert space description, 
so there is a possibility that we can completely write down the theory in the small Hilbert space description, but it is not yet clear. 
In the next section, we will see some results which are yielded from this action.

\section{Some Results} \label{5}
In this section, we study some properties of type II string field theory which was proposed in the last section. 
In paticular, we would like to check whether our action can reproduce correct amplitudes expected from the first quantized theory.

\subsection{Perturbation Theory: the Kinetic Term}

In the last section, we proposed a concrete form of the type II full action $S$ via bosonic pure gauge solutions as functionals of type II string fields. 
The full action $S$ produces the free term $S_{2}$ and every interaction term $S_{3},S_{4},\dots $ in each order of $\kappa$ through the following expansion: 
\begin{align}
S=\sum_{n=2}^{\infty} \kappa ^{n-2} S_{n} .
\end{align}
In this subsection, we pick up the kinetic term and see details: the gauge transformation and the cohomology, or the physical state space. 
In the next subsection, we will see the properties of lower-order terms. 


\vspace{3mm}

A type II string field $\Psi$ satisfies the level matching condition $L ^{-}_{0} \Psi =0$ ($L ^{-}_{0} := L_{0} - \bar{L}_{0}$) as other closed string fields. 
The free type II action in the NS-NS sector is given by 
\begin{align}
S_{2}= \int_{0}^{1}{dt} \, \langle \partial _{t}(t\Psi ), Q (t\Psi )\rangle _{\eta } 
= \frac{1}{2}\langle \eta ^{+}_{0}\eta ^{-}_{0} \Psi , Q \Psi \rangle ,
\end{align}
where $Q$ is the BRST operator of type II superstrings and we define $\eta ^{\pm }_{0}:=\eta _{0}\pm \bar{\eta }_{0}$.
The inner product is defined as $\langle A,B\rangle =\langle A|c_{0}^{-}|B\rangle$. 
Here, $\langle A|$ is the BPZ conjugation of $|A\rangle$. 
This inner product is nondegenerate just on $\mathrm{Ker}[b_{0}^{-}]$.
The normalization of correlators in the full conformal field theory for a flat space-time background is given by 
\begin{align}
\langle \xi (w_{1})\bar{\xi }(\bar{w}_{1})e^{-2\phi (w_{2})}e^{-2\bar{\phi }(\bar{w}_{2})} & c(z_{1})\bar{c}(\bar{z}_{1})c(z_{2})\bar{c}(\bar{z}_{2})c(z_{3})\bar{c}(\bar{z}_{3})e^{ip_{\mu}X^{\mu}(z,\bar{z})}\rangle 
\\ \nonumber 
&=2(2\pi )^{D}\delta ^{D}(p)|z_{1}-z_{2}|^{2}|z_{1}-z_{3}|^{2}|z_{2}-z_{3}|^{2},
\end{align}
where $D=10$ is the critical dimension of space-time. Since we use the $c_{0}^{-}$-inserted inner product, the total ghost and picture number in the action must be $(g,p)=(3,-2)$. The ghost and picture number $(g,p)$ of $\Psi $ are $(g,p)=(0,0)$, so if necessary, we write these explicitly as $\Psi \equiv \Psi _{(0,0)}$.

This free action is invariant under the gauge transformation $\Psi \mapsto \Psi + Q \Lambda $ in $\mathcal{H}_{\eta }$. 
In the large Hilbert space description, it is equivalent to the following gauge transformation:
\begin{align}
\delta \Psi =Q\Lambda +\eta ^{+}_{0}\Omega^{+} +\eta ^{-}_{0}\Omega^{-} ,
\end{align}
where $\Lambda =\Lambda _{(-1,0)}$, $\Omega ^{+}=\Omega ^{+}_{(-1,1)}$, and $\Omega ^{-} =\Omega ^{-}_{(-1,1)}$ all are gauge parameters. 
Then the equation of motion of free theory is given as follows, 
\begin{align}
Q \eta ^{+}_{0}\eta ^{-}_{0} \Psi =0 \hspace{5mm} (\Psi \in \mathcal{H}_{L}) \, .
\end{align}
Using the identification $\eta _{0}^{+} \eta _{0}^{-} \Psi =\Psi _{S}$, it is equivalent to the ordinary small Hilbert space one:  
\begin{align}
Q\Psi _{S}=0  \hspace{5mm} (\Psi _{S} \in \mathcal{H}_{S}) \, ,
\end{align}
where $\Psi_{S}$ is the type II string field in the small Hilbert space description 
and $\xi^{\pm }_{0}:=\frac{1}{2}(\xi _{0} \pm \bar{\xi }_{0})$. 
In the quotient space $\mathcal{H}_{\eta }$, 
choosing a representative element, we can write $\Psi \in \mathcal{H}_{\eta }$ as 
$\Psi \overset{\mathrm{eqv.}}{=}[ \xi_{0}^{-}\xi _{0}^{+} \hat{\psi} ]$ where $\hat{\psi}_{(2,-2)} \in \mathrm{Ker}[\eta^{\pm }_{0}]$. 
So the equation of motion becomes 
\begin{align}
 Q \Psi 
&\stackrel{\mathrm{eqv.}}{=}  \left[ \xi ^{-}_{0} \xi ^{+}_{0} Q \hat{\psi} _{(2,-2)} \right] = 0 \hspace{5mm } (\Psi \in \mathcal{H}_{\eta } ) \, 
\end{align}
and we obtain $Q \hat{\psi}_{(2,-2)}=0$, which is equivalent to above two equations. 
They all give the condition of the first quantization. 
It implies that physical state spaces which are described in the large/small Hilbert space $\mathcal{H}_{L}$/$\mathcal{H}_{S}$ and the quotient space  $\mathcal{H}_{\eta }$ are equavalent to each other,
\begin{align}
\frac{\mathrm{Ker}[Q:\mathcal{H}_{\eta }\rightarrow \mathcal{H}_{\eta }]}
{\mathrm{Im}[Q:\mathcal{H}_{\eta }\rightarrow \mathcal{H}_{\eta }]}
\cong 
\frac{\mathrm{Ker}[\eta^{+}_{0}\eta^{-}_{0}Q:\mathcal{H}_{L}\rightarrow \mathcal{H}_{L}]}
{\mathrm{Im}[Q,\eta^{\pm }:\mathcal{H}_{L}\rightarrow \mathcal{H}_{L}]}
\cong 
\frac{\mathrm{Ker}[Q:\mathcal{H}_{S}\rightarrow \mathcal{H}_{S}]}
{\mathrm{Im}[Q:\mathcal{H}_{S}\rightarrow \mathcal{H}_{S}]} .
\end{align}

\subsection{Perturbation Theory: Lower-order Interacting Terms}


In the last section, the type II full action was given by the following form:
\begin{align}
S
=\int_{0}^{1}{dt} \, \langle \partial _{t} \Psi (t) , Q_{\mathcal{G}}
 \Psi (t) \rangle _{\eta }
\equiv \int_{0}^{1}{dt} \, \sum_{n=0}^{\infty }  \frac{\kappa ^{n}}{n!}  \langle \partial _{t} \Psi (t) , [ \mathcal{G}(t)^{n} , \Psi (t) ] \rangle _{\eta } .
\end{align}
It would reproduce all interaction terms of type II string field theory. 
In this subsection, we see the result yielded from the lower-order terms $S_{2}+\kappa S_{3}+\kappa ^{2}S_{4}+O(\kappa ^{3})$. 
Recall that the pure gauge solution $\mathcal{G}(t)$ is expanded with respect to the coupling constant $\kappa$ as follows: 
\begin{align}
\mathcal{G}(\Psi )&= Q\mathcal{P}_{-}Q\Psi 
+ \frac{\kappa }{2}\Big(  \big[ Q\mathcal{P}_{-}Q\Psi , \mathcal{P}_{-}Q\Psi \big] + Q\mathcal{P}_{-} \big[ Q\mathcal{P}_{-}Q\Psi , \Psi \big]  \Big)
\\ \nonumber
& \hspace{4mm} +\frac{\kappa ^{2}}{3!} \Big(  \big[ \big( Q\mathcal{P}_{-}Q\Psi \big) ^{2}, \mathcal{P}_{-} Q \Psi \big] + \big[  [ Q\mathcal{P}_{-}Q\Psi , \mathcal{P}_{-}Q\Psi ] +  Q\mathcal{P}_{-}[ Q\mathcal{P}_{-}Q\Psi  , \Psi  ]  , \mathcal{P}_{-}Q\Psi \big] 
\\ \nonumber
& \hspace{5mm} 
+ Q\mathcal{P}_{-} \Big(  \big[ \big( Q\mathcal{P}_{-}Q\Psi \big) ^{2}, \Psi \big] + \big[  [ Q\mathcal{P}_{-}Q\Psi , \mathcal{P}_{-}Q\Psi ] + Q\mathcal{P}_{-}[Q\mathcal{P}_{-}Q\Psi , \Psi ]  , \Psi \big]  \Big) \Big)
+\dots .
\end{align}
Therefore we obtain the following lower-order perturbative action:
\begin{align}
\nonumber 
S&= \int_{0}^{1}{dt} \, \Big( \langle \partial _{t} \Psi (t) , Q \Psi (t) + \frac{\kappa }{2}[ \mathcal{G}(t), \Psi (t) ] +\frac{\kappa ^{2}}{3!}[\mathcal{G}(t)^{2} , \Psi (t) ] \rangle _{\eta } + \dots \Big)  
\\ \nonumber 
&= \int_{0}^{1}{dt} \, \langle \partial _{t} \Psi , Q\Psi \rangle _{\eta } 
+ \kappa \int_{0}^{1}{dt} \, \langle \Psi , [ Q\mathcal{P}_{-}Q\Psi , \Psi ] \rangle _{\eta }
+\frac{\kappa ^{2}}{2} \int_{0}^{1}{dt} \, \langle \partial _{t} \Psi , [ (Q\mathcal{P}Q\Psi )^{2} , \Psi ] \rangle _{\eta } 
\\
& \hspace{6mm} + \frac{\kappa ^{2}}{ 2} \int_{0}^{1}{dt} \, \langle  \Psi , \big[ [Q\mathcal{P}_{-}Q\Psi , \mathcal{P}_{-}Q\Psi ] + Q\mathcal{P}_{-}[Q\mathcal{P}_{-}Q\Psi , \Psi ] , \Psi \big] \rangle _{\eta } +\dots .
\end{align}
Then, the variation of the action becomes the following form: 
\begin{align}
\nonumber 
\delta S &= \int_{0}^{1}{dt} \, \partial _{t} \langle \delta \Psi (t) , Q_{\mathcal{G}(t)}\Psi (t) \rangle _{\eta } 
= \langle \delta \Psi , Q_{\mathcal{G}} \Psi \rangle _{\eta }
\\ \nonumber 
&= \langle \delta \Psi ,  Q \Psi \rangle _{\eta }
+ \kappa \langle \delta \Psi , [ Q\mathcal{P}_{-}Q\Psi , \Psi ] \rangle _{\eta }
+  \frac{\kappa ^{2}}{2} \langle \delta \Psi , \big[ (Q\mathcal{P}_{-}Q\Psi )^{2}  , \Psi \big] \rangle _{\eta }
\\
& \hspace{4mm} + \frac{\kappa ^{2}}{2} \langle  \delta \Psi , \big[  [ Q\mathcal{P}_{-}Q\Psi , \mathcal{P}_{-}Q\Psi ] +Q\mathcal{P}_{-}[ Q\mathcal{P}_{-}Q\Psi , \Psi ] , \Psi \big] \rangle _{\eta }
+ \dots .
\end{align}
The gauge transformation is therefore given as follows: 
\begin{align}
\nonumber 
\delta \Psi  
= & Q\Lambda + \kappa  \big[ Q\mathcal{P}_{-}Q\Psi , \Lambda \big] +\frac{\kappa ^{2}}{2}  \big[ (Q\mathcal{P}_{-}Q\Psi )^{2},\Lambda \big] 
\\ 
&+ \frac{\kappa ^{2}}{2}\Big( \big[ [ Q\mathcal{P}_{-}Q\Psi ,\mathcal{P}_{-}Q \Psi ] + Q\mathcal{P}_{-}[ Q\mathcal{P}_{-}Q\Psi , \Psi ] , \Lambda \big] \Big) + \dots .
\end{align}
Let us check the gauge invariance of the action perturbatively. 
We can write it down by the sum of the following all lines which are equal to zero as follows: 
\begin{align}
\nonumber 
 \kappa ^{0}:& \,\,\,\,\, \hspace{40mm}  \langle Q\Lambda ,  Q \Psi \rangle _{\eta } =0    
\\ \nonumber 
 \kappa ^{1}:& \,\,\,\,\,  \hspace{10mm}  \langle Q\Lambda ,  \big[ Q\mathcal{P}_{-}Q\Psi , \Psi \big] \rangle _{\eta } 
+ \langle \big[ Q\mathcal{P}_{-}Q\Psi , \Lambda \big] , Q \Psi \rangle _{\eta } =0   
\\ \nonumber 
 \kappa ^{2}:& \,\,\,\,\,  \langle Q\Lambda , \big[ (Q\mathcal{P}_{-}Q\Psi )^{2} , \Psi  \big] 
+ \big[ [ Q\mathcal{P}_{-}Q\Psi , \mathcal{P}_{-}Q\Psi ]
+  Q\mathcal{P}_{-}[Q\mathcal{P}_{-}Q\Psi , \Psi ] , \Psi \big] 
 \rangle _{\eta }
 \\ \nonumber 
& \hspace{4mm} + \langle  [ ( Q \mathcal{P}_{-} Q \Psi )^{2} , \Lambda ] 
+ \big[ [ Q \mathcal{P}_{-} Q \Psi , \mathcal{P}_{-} Q \Psi ] 
+  Q \mathcal{P}_{-} [ Q \mathcal{P}_{-} Q \Psi , \Psi ]  , \Lambda \big] , Q \Psi \rangle _{\eta }
\\ \nonumber 
& \hspace{6mm} +  \langle [Q\mathcal{P}_{-}Q\Psi , \Lambda ] , [ Q\mathcal{P}_{-}Q \Psi , \Psi  ] \rangle _{\eta }
+ \langle [Q\mathcal{P}_{-}Q\Psi , \Lambda ] , [ Q\mathcal{P}_{-}Q \Psi , \Psi  ] \rangle _{\eta } 
=0
\\
 \kappa ^{n}:& \,\,\,\,\,  \hspace{50mm} \vdots
\end{align}
Vanishing of the first line is equivalent to the nilpotency of the BRST operator: $Q^{2}=0$, 
that of the second line means the derivation propertiy: $Q[A,B]+[QA,B]+(-1)^{A}[A,QB]=0$, 
and that of the third line implies the lowest-order (homotopy-) Jacobi identity: $\Delta _{Q}[A,B,C]+\Delta _{\mathcal{J}}[A,B,C]=0$ where 
\begin{align}
\Delta _{Q}[A,B,C] &= Q[A,B,C] + [QA,B,C]+(-1)^{A}[A,QB,C]+(-1)^{A+B}[A,B,QC] , 
\\
\Delta _{\mathcal{J}}[A,B,C] &= [A,[B,C]]+(-1)^{A(B+C)}[B,[C,A]]+(-1)^{C(A+B)}[C,[A,B]] .
\end{align}
We can also read higher-order relations from vanishing of higher-order terms similarly: $\Delta _{Q}[A_{1},\dots ,A_{n}]+\Delta _{\mathcal{J}}[A_{1},\dots ,A_{n}]=0$, where $\Delta _{Q/\mathcal{J}}[A_{1},\dots ,A_{n}]$ means the violation of the derivation/Jacobi property of the $n$-product $[A_{1},\dots ,A_{n}]$. 

\subsection{Three Point Amplitudes}

Let us consider three point amplitudes. 
Recall that the $\kappa ^{1}$-order action is given by
\begin{align}
S_{3}= \int_{0}^{1}{dt} \langle \partial _{t} \Psi (t), [\Psi (t), Q\mathcal{P}_{-}Q\Psi (t)] \rangle _{\eta } 
 =\frac{1}{3} \langle 
\Psi , [ \Psi , Q \mathcal{P}_{-} Q \Psi ] \rangle _{\eta } ,
\end{align}
where $\Psi (t) = t\Psi$. 
This cubic term of the action reduces to the expected correlator for physical states. 
Writing $\Psi =\xi _{0}^{+}\xi _{0}^{-} \hat{\Psi }_{(2,-2)} $ where $Q\hat{\Psi }_{(2,-2)}=0$ and $\mathcal{P}_{-}=\eta _{0}^{-}\xi_{0}^{-}$, we obtain 
\begin{align}
Q\mathcal{P}_{-}Q\Psi = Q\eta _{0}^{-} \xi _{0}^{-} Q \xi _{0}^{+}\xi _{0}^{-} \hat{\Psi }_{(2,-2)}
=Q\mathcal{X}_{0}^{-}\xi_{0}^{+}\hat{\Psi }_{(2,-2)} 
\overset{e.o.m.}{=}  \mathcal{X}_{0}^{+}\mathcal{X}_{0}^{-}\hat{\Psi }_{(2,-2)}. 
\end{align}
Note that the last equal sign holds on the mass shell. 
Hence, we see that
\begin{align}
S_{3} = \frac{1}{3} \langle \hat{\Psi },  [ \xi_{0}^{+}\xi_{0}^{-}\hat{\Psi} , \mathcal{X}_{0}^{+}\mathcal{X}_{0}^{-}\hat{\Psi }  ]  \rangle \cong  
\frac{1}{3} \langle \langle \hat{\Psi } , \hat{\Psi } , \mathcal{X}_{0}^{+}\mathcal{X}_{0}^{-}\hat{\Psi } \rangle \rangle .
\end{align}
Here, the symbol $\langle \langle A, B ,C \rangle \rangle$ denotes the correlator. 
This is the expected result. 

\vspace{2mm}

{\parindent=0pt \underline{Pure Gauge Solutions in the Correlator}}

\vspace{1mm}

In general, there is the correspondence between correlation functions in the small Hilbert space $\langle \langle A_{1}, \dots , A_{n} \rangle \rangle$ and those in the large Hilbert space, where $\phi$ and $\bar{\phi }$ are bosonized superghosts:  
\begin{align}
 \langle \langle A_{1}, \dots ,A_{n} \rangle \rangle  =  \langle \xi \bar{\xi } e^{-2\phi}e^{-2\bar{\phi }} A_{1} \dots A_{n} \rangle .
\end{align}
This implies that the correlator selects one $\xi^{+}_{0}\xi^{-}_{0}$-component and other small space components. 
For example, suppose $\{ A_{i} \} _{i=1}^{n}$ belong to the small Hilbert space, and $\Phi$/$\Psi $ can be $\xi _{0}^{\pm }$-decomposed as $\Phi = \hat{\phi }+ \xi _{0}^{+}\xi _{0}^{-}\hat{\phi}_{\xi }$/$\Psi = \hat{\psi }+ \xi _{0}^{+}\xi _{0}^{-}\hat{\psi}_{\xi }$, then the correlation function of these elements is given by the following form: 
\begin{align}
\langle \eta _{0}^{+} \eta _{0}^{-}\Phi ,[ \Psi , A_{1}, \dots , A_{n} ] \rangle = \langle \hat{\phi }_{\xi } , [ \xi_{0}^{+}\xi_{0}^{-} \hat{\psi }_{\xi } ,A_{1}, \dots ,A_{n} ] \rangle .
\end{align}

Our type II action was given by this form: 
\begin{align}
S= \int_{0}^{1}{dt} \, \langle  \partial _{t} \Psi , Q_{\mathcal{G}} \Psi \rangle _{\eta } 
=\int_{0}^{1}{dt} \sum_{n=0}^{\infty } \frac{\kappa ^{n}}{n!} \langle \eta _{0}^{+} \eta _{0}^{-} \Psi , [ \mathcal{G}(t\Psi )^{n} , t\Psi ] \rangle .
\end{align}
In the next subsection, we estimate higher-point amplitudes. 
To this purpose, in the rest of this subsection, we see the property of $\mathcal{G}$ in the correlator. 
Note that using (\ref{XG}) for $\eta _{0}^{\pm }$ and $\eta _{0}^{+} \eta _{0}^{-} \mathcal{G}(\Psi ) =0 $, we obtain the $\xi_{0}^{\pm }$-decomposition of $\mathcal{G}(\Psi )$:
\begin{align}
\mathcal{G}(\Psi ) = \hat{\psi }_{Q} - \xi ^{+}_{0} \big( Q_{\mathcal{G}} \Lambda _{\eta ^{+}} \big)  - \xi ^{-}_{0} \big(  Q_{\mathcal{G}} \Lambda _{\eta ^{-}} \big) .
\end{align}
In our construction, it becomes $\mathcal{G}(\Psi )=\hat{\psi}_{Q}-\xi _{0}^{+}Q_{\mathcal{G} }\Lambda _{\eta }$ because $\eta ^{-}_{0} \mathcal{G}(\Psi )=0$. 
Then we notice that $\xi _{0}^{\pm }$-components of $\mathcal{G}(\Psi )$ are given by BRST-exact terms. 
As a result, it is expected that the value of the correlation function depends only on the small space component of $\mathcal{G}(\Psi )$. 
We therefore regard the pure gauge solution $\mathcal{G}(\Psi )$ as an element of the small Hilbert space: 
\begin{align}
\mathcal{G}(\Psi ) \cong   {P}_{+}\mathcal{P}_{-} \mathcal{G}(\Psi ) = \hat{\psi }_{Q} ,
\end{align}
although it is constructed from type II string fields $\Psi$ which belong to the large Hilbert space. 

Since there are one $\eta _{0}^{+}$ and one $\eta _{0}^{-}$ in our large Hilbert space action, 
the value of correlation functions in our theory will be determinded by the large Hilbert space component of $\Psi$ and the small Hilbert space component of $\mathcal{G}(\Psi )$. 

\subsection{Reduction to Small Hilbert Space}
Let us consider amplitudes, which are expected to correspond to the result of the first quatization theory. 
To this purpose, the small Hilbert space description is rather reasonable because a small string field $\Psi _{S}$ naturaly corresponds to the vertex operator of closed strings. 
We therefore consider the reduction to the small Hilbert space, and seek fundamental vertices which produce Feynman graphs of string interaction in the small Hilbert space. 
We would like to expect that there exists the correspondence $\Psi _{S}\equiv \eta _{0}^{+}\eta _{0}^{-} \Psi $ between the small string field $\Psi _{S}$ and the large string field $\Psi $ as the free case. 
To see it explicitly, let us consider the zero mode decomposition: 
\begin{align}
\Psi = \hat{\psi }^{\circ }+\xi ^{+}_{0}\hat{\psi }^{+} +\xi ^{-}_{0}\hat{\psi }^{-} +\xi^{+}_{0} \xi^{-}_{0} \hat{\psi }  
\end{align}
where $\hat{\psi}^{\circ }$, $\hat{\psi }^{\pm}$, and $\hat{\psi }$ belong to $\mathrm{Ker}[\eta^{\pm }_{0}]$, namely, the small Hilbert space $\mathcal{H}_{S}$. 
Then the large component $\hat{\psi }$ has ghost number $2$ and picture nummber $-2$. 
(We often write it explicitly as $\hat{\psi } \equiv \hat{\psi }_{(2,-2)}$.) 
Therefore, we identify the small string field $\Psi _{S}$ with this $\hat{\psi }$ up to a sign factor: 
\begin{align}
\Psi _{S}  \Longleftrightarrow  - \hat{\psi }_{(2,-2)} .
\end{align}
It is equivalent to impose on $\Psi$ the following (partial) fixing conditions:
\begin{align}
\xi ^{+}_{0} \Psi = \xi ^{-}_{0} \Psi = 0 .
\end{align}
We assume that the other elements $\hat{\psi }^{\circ }$ and $\hat{\psi }^{\pm }$ correspond to auxiliary fields as well as the correspondence in free theory. 
We thus consentrate on the part which is related to the element $\hat{\psi }_{(2,-2)}$. 
The pure gauge solution, which is a functional of type II string fields, is given by
\begin{align}
\mathcal{G}(\Psi ) =Q\mathcal{P}_{-}Q \Psi + \frac{\kappa }{2} \Big( [Q\mathcal{P}_{-}Q \Psi  ,   \mathcal{P}_{-}Q \Psi ] + Q\mathcal{P}_{-}[ Q\mathcal{P}_{-}Q \Psi , \Psi ] \Big)  
+ \dots  .
\end{align}
We can calculate the $\xi _{0}^{\pm }$-decomposition of $Q\mathcal{P}_{-}Q\Psi$, $\mathcal{P}_{-}Q\Psi$, and $Q\mathcal{P}_{-}\Psi$ as follows: 
\begin{align}
\nonumber 
Q\mathcal{P}_{-}Q\Psi  \, \cong 
\mathcal{P}_{+}Q\mathcal{P}_{-}Q\Psi \, =
 \, & \mathcal{P}_{+}Q\mathcal{P}_{-} Q \left( \hat{\psi }^{\circ }+ \xi^{+}_{0}\hat{\psi }^{+} +\xi^{-}_{0}\hat{\psi }^{-} +\xi^{+}_{0} \xi^{-}_{0}\hat{\psi }  \right) 
\\ \nonumber 
= \, & \mathcal{P}_{+}Q\left( Q\hat{\psi }^{\circ  }+ Q\xi^{+}_{0}\hat{\psi }^{+} + \mathcal{X}^{-}_{0}\hat{\psi }^{-}  - \mathcal{X}^{-}_{0} \xi^{+}_{0}\hat{\psi } \right) 
\\ 
= \, & Q \mathcal{X}^{+}_{0}\hat{\psi }^{+} + Q \mathcal{X}^{-}_{0}\hat{\psi }^{-} -  \mathcal{X}^{+}_{0} \mathcal{X}^{-}_{0} \hat{\psi } ,
\\ \nonumber 
\mathcal{P}_{-}Q\Psi  \, \cong 
\mathcal{P}_{+}\mathcal{P}_{-}Q\Psi \, =
\, & \mathcal{P}_{+}\mathcal{P}_{-} Q \left( \hat{\psi }^{\circ } +\xi^{+}_{0}\hat{\psi }^{+} +\xi^{-}_{0}\hat{\psi }^{-} +\xi^{+}_{0} \xi^{-}_{0}\hat{\psi }  \right) 
\\ 
= \, &  Q\hat{\psi }^{\circ }  + \mathcal{X}^{+}_{0}\hat{\psi }^{+} + \mathcal{X}^{-}_{0}\hat{\psi }^{-} ,
\\
Q\mathcal{P}_{-}\Psi \, \cong 
\mathcal{P}_{+}Q\mathcal{P}_{-} \Psi \, = 
\, & Q\hat{\psi }^{\circ } +\mathcal{X}_{0}^{+} \hat{\psi }^{+} .
\end{align}
Suppose auxiliary fields are integrated out and BRST-exact terms are dropped, then we obtain   
\begin{align}
Q\mathcal{P}_{-}\Psi =\mathcal{P}_{-}Q\Psi  \,\, = \, 0   \,\, , 
\hspace{5mm} Q\mathcal{P}_{-}Q\Psi  \,\, = \, -\mathcal{X}^{+}_{0}\mathcal{X}^{-}_{0}\hat{\psi } \,\, .
\end{align}
Thus we regard $\mathcal{G}(\Psi ) \cong  -\mathcal{X}_{0}^{+} \mathcal{X}_{0}^{-} \hat{\psi }$ in the small Hilbert space description. 
Using the identification $\Psi _{S} \cong  -\hat{\psi }_{(2,-2)}$, 
we can write down $\mathcal{F}(\mathcal{G}(\Psi ))=\mathcal{F} \left( \mathcal{X}_{0}^{+} \mathcal{X}_{0}^{-} \Psi _{S}  \right) = 0$ as follows:
\begin{align}
Q \left(  \mathcal{X}_{0}^{+} \mathcal{X}_{0}^{-} \Psi _{S}  \right) + \sum_{n=1}^{\infty} \frac{\kappa ^{n}}{(n+1)! }[ \left( \mathcal{X}_{0}^{+} \mathcal{X}_{0}^{-} \Psi _{S} \right) ^{n}  , \mathcal{X}_{0}^{+} \mathcal{X}_{0}^{-} \Psi _{S}  ]
\cong 0 .
\end{align}
Then we can estimate the following part of the small Hilbert space action, which is expected to contribute to the value of correlation functions: 
\begin{align}
\widetilde{S}= \frac{1}{2} \langle \langle \Psi _{S} , Q \Psi _{S} \rangle \rangle + \sum_{n=1}^{\infty }\sum_{\sigma }  \frac{\kappa ^{n}}{(n+2)!} \langle \langle \Psi _{S} , \sigma \big(  \big[ ( \mathcal{X}_{0}^{+} \mathcal{X}_{0}^{-} \Psi _{S} ) ^{n} ,\Psi _{S} \big] \big)  \rangle \rangle ,
\end{align}
where $\sigma $ is a permutation and the sum of $\sigma $ runs over all possible permutations. 
Here, $\langle \langle A, B\rangle \rangle $ is the inner product in the small Hilbert space. 
It is nonzero if and only if the pair of the total ghost and picture nummber is equal to $(g,p)=(5,4)$. 


Now, let us consider string fundamental vertices, or equivalentry, amplitudes. 
We would like to read small space $n$-point vertices $\widetilde{S}_{n}$ of $\kappa ^{2}\widetilde{S}=\sum \kappa ^{n} \widetilde{S}_{n}$ from the large space action $\kappa ^{2}S=\sum \kappa ^{n}S_{n}$. 
First, we consider the three point vertex. 
In this case, we can read it quickly as 
\begin{align}
S_{3} = \frac{1}{3}
\langle 
\Psi , [ \Psi ,Q \mathcal{P}_{-}Q \Psi ] \rangle 
_{\eta }
\longrightarrow \,\,
\widetilde{S}_{3}= \frac{1}{3} \{ \Psi _{S} , 
 \Psi _{S} , \mathcal{X}^{+}_{0} \mathcal{X}^{-}_{0} \Psi _{S} \} _{S} 
\end{align}
where $\{ A, B ,C \} _{S} := \langle \langle A , [ B ,C ] \rangle \rangle $. 
It is an expected one. 
Next, we considr higher-point vertices. 
Recall that the $\kappa ^{n}$-order large space action is given by 
\begin{align}
S_{n} = \frac{1}{n!}  \sum_{\sigma }  \langle 
\Psi , \, \sigma \Big( \big[ ( Q\mathcal{P}_{-}Q\Psi )^{n-2} , \Psi \big] +\dots + \big[ Q\mathcal{P}_{-}[\dots [\Psi, Q\mathcal{P}_{-}Q\Psi ] ],\Psi \big] \Big) \rangle _{\eta } .
\end{align}
From the correspondence discussed above, we can estimate higher-point vertices similarly: 
\begin{align}
\widetilde{S}_{n} 
= \frac{1}{n!} \sum_{\sigma } \{ \Psi _{S} , \, \sigma \big( \Psi _{S} , \, ( \mathcal{X}^{+}_{0} \mathcal{X}^{-}_{0} \Psi _{S} )^{n-2} \big) \} _{S} .
\end{align}
This is a result what we expected from the correspondence to the first quantization theory. 
Since we use bosonic closed string products, the single covering of the moduli space of Riemann surfaces is realized automatically. 
The problem of the divergence with respect to the local picture changing operators would be also resolved by the nilpotency of homotopy algebras.


\section{Summary and Discussion} \label{6}

In this paper, using algebraic properties of closed strings, we gave a concrete construction of type II string field theory. 
It is based on the large Hilbert space description like \cite{Berkovits:1995ab, Okawa:2004ii}, so there are no local insertions of picture changing operators. 
In the construction of supersymmetric theory, a $formal$ pure gauge solution of bosonic theory has played a nonnegligible role. 
A pure gauge solution of bosonic string field theory is always constructed from gauge parameter fields \cite{Schubert:1991en}. 
In the large Hilbert space description, we have constructed superstring field theories by identifying superstring fields with these gauge parameter fields, so it is with our construction. 
Making a pure gauge solution as a functional of type II string fields, we construct the full action for type II string field theory. 
We would like to note that our construction is based on algebraic properties of closed string fields, $L_{\infty }$-algebras. 
It is a result from that we use the extension of Zwiebach's closed string products\cite{Zwiebach:1992ie}. 
(Recall that the action for closed string field theory whose Feynman graphs reproduce a single covering of the moduli space has the geometrical vertices satisfying the algebraic relations of the BV-master equation, $L_{\infty }$-algebras and their quantum extensions \cite{Stasheff:1993ny, Kimura:1993ea, Kajiura:2001ng}.) 
Thus, in this paper, we did not touch the geometrical understanding of superstrings like the correspondence of the full action and the decomposition of the moduli space of super-Riemann surfaces as the case of bosonic theory. 
In the rest of this section, we would like to disucuss related aspects of this type II action and put some comments. 

\vspace{2mm}

{\parindent=0pt { \underline{Comparison with Bosonic Theory} }}

\vspace{1mm}

We can rewrite the bosonic action into $\int{dt}\langle \partial \Psi , Q_{\Psi }^{\prime }\Psi \rangle $, which consits of the Maurer-Cartan operator $Q_{\Psi}^{\prime }$ and has the gauge invariance generated by the BRST operator $Q_{\Psi }$. 
On the other hand, we  write down the type II action by $\int{dt}\langle \partial \Psi , Q_{\mathcal{G}}\Psi \rangle _{\eta }$, which consists of the BRST operator $Q_{\mathcal{G}}$ and has the gauge invariance generated by the BRST operator $Q_{\mathcal{G}}$. 
In bosonic theory, we can construct the Maurer-Cartan operator $Q^{\prime }_{\Psi }$ as the linear combination of the BRST operators $Q_{\Psi }^{[a]}$. 
We regard this $Q_{\Psi }^{[a]}$ as a deformation of the gauge structure because $Q_{\Psi}^{[0]}$ generates the gauge transformation of free theory and $Q_{\Psi }^{[1]}$ generates that of full theory. 
In type II theory, we can introduce $Q^{[a]}_{\mathcal{G}}$ by the replacement of $\kappa \rightarrow a\kappa$ and define a Maurer-Cartan (like) operator $Q_{\mathcal{G}}^{\prime }$ by the integration of $a$ from $0$ to $1$. 
Using this operator, we can write down the another action as $S^{\prime }=\int{dt}\langle \partial \Psi , Q_{\mathcal{G}}^{\prime }\Psi \rangle _{\eta }$, which gives $Q^{\prime }_{\mathcal{G}}\Psi =0$ as the equation of motion. 
Then the perturbative expansion of $\kappa ^{2}S^{\prime }=\sum{\kappa ^{n} S_{n}^{\prime } }$ gives the almost same result as 
our action $\kappa ^{2}S=\sum{ \kappa ^{n} S_{n} }$. 
The difference is the coefficients of $n$-point vertices: $(n-1)S_{n}^{\prime }= \mathcal{S}_{n}$. 
Note that this $S^{\prime }$ is not invariant under the transformation generated by $Q_{\mathcal{G}}$ (but $Q_{\mathcal{G}}^{[a]}$).\footnote{
Of course, there is some possibilities that the gauge transformatin is realized by the nontrivial form as Berkovits theory: the element generating gauge symmetry is not $\delta \Psi$ but some combination of $\Psi$ like $\delta (e^{\Phi })$. 
} 

\vspace{2mm}

{\parindent=0pt{ \underline{Pure Gauge Solution and $L_{\infty }$-structure}  }}

\vspace{1mm}

In our construction, we used the method which is based only on the properties of nilpotent homotopy algebras, so this framework is available for considering other gauge theories as long as the gauge invariance is governed by nilpotent homotopy algebras. 
In particular, considering a shift by a pure gauge solution $\mathcal{G}$, we wrote down the type II action. 
However, we would like to note that there exist some ambiguities in constructing a pure gauge solution. 
For example, we can use some another projector like $\mathcal{P}_{+}$, or modify the defining equation of the pure gauge solution as $\partial _{\tau } \mathcal{G}=Q_{\mathcal{G}}\mathcal{P}_{-}Q\Psi$, and so on. 
Or as well as open string field theory, in closed string field theory, we may be able to construct a pure gauge solution without using a differential (or integral) equation which we used. 
In this article, we chose simplest one which realizes $\delta \mathcal{G}(\Psi )=Q_{\mathcal{G}}\Lambda _{\delta }=0$ under the gauge transformation because we would $Q_{\mathcal{G}}$'s expanding point $\mathcal{G}$ like to belong the equivalent class of the gauge transformation.

\vspace{2mm}

{\parindent=0pt{ \underline{Algebraic structure of Type II String Field Theory} } }

\vspace{1mm}

The gauge symmetry of string field theory is infinitly reducible. 
Thus it is not clear whether we can obtain a gauge fixed action as a simple extension of the original action \cite{Kroyter:2012ni, Torii:2012nj, Kohriki:2012}.
Since the type II action consists of string fields, the BRST operator $Q$, a projector $\mathcal{P}_{-}$, and closed string products, 
we can expect that guage fixing is carrid out by relaxing the constraint of the ghost number of string fields as the case of bosonic theory. 
However, to prove it exactly, we need to construct vertices for type II fields (not for bosonic fields) from the decomposition of the moduli space of super-Riemann surfaces, which would naturally lead to (extended) homotopy algebras of type II fields. 
We need these geometrical understandings of superstring fields for quantization, constructing theory which includes $R$-sector, considering the problem of background independence, other nonperturbative effects, and so on. 
It is rather not clear in the case of the Large Hilbert space description. 
In particular, in the large Hilbert space description, we know the cyclicity of a graded symplectic form and the sign of products of homotopy algebras are not well-suited naively, so we also need to find a good algebraic structure of supersting fields on this point. 
Recently, there is a progress in this geometrical point of view \cite{Jurco:2013qra}. 
They gave an outline to construct type II vertices and the quantum BV-master action geometrically in the Small Hilbert space description. 
In quantum bosonic closed string field theory, the vertices satisfy the relations of a loop homotopy Lie algebra
, whose classical part gives $L_{\infty }$-algebras. 
In \cite{Jurco:2013qra}, they introduced $\mathcal{N}=1$ loop homotopy Lie algebras by considering the operad which deeply relates to type II string field theory.

\vspace{3mm}

{\parindent=0pt{ {\bf{Ackonledgements:}} }}

This is an extended work of my master's thesis, so I would like to thank Yuji Okawa and Mitsuhiro Kato, my supervisors, for fruitful discussions, suggestive comments, and their kindness. 
I am also grateful to the members of our Komaba particle theory group, in particular, Yuki Iimori, Shota Komatsu, and Shingo Torii for useful discussions and comments, and their kindness. 
I would also like to thank the organizers of the conference `String Field Theory and Related Aspects: SFT 2012' hosted by the Israel Institute for Advanced Studies. 


\appendix

\section{$L_{\infty }$-algebras and Closed String Field Theory} \label{Appendix A}

A representation of the $L_{\infty }$-operad on a fixed graded vector space $\mathcal{H}$ is an $L_{\infty }$-algebra $(\mathcal{H},{\bf L})$ 
\cite{Stasheff:1993ny, Kimura:1993ea, Kajiura:2005sn, Getzler:2007}. 
In string field theory, we regard the string state space, Hilbert space of conformal field theory, as a graded vector space $\mathcal{H}$ through the identification of the world sheet ghost number and $\mathcal{H}$'s grading 
\cite{Schwarz:1992nx, Sen:1993, Hata:1993gf, Gaberdiel:1997ia, Stasheff:1993ny, Kimura:1993ea, Kontsevich:1997vb, Nakatsu:2001da,
Kajiura:2001ng}. 
In Appendix A, we give a short review of the $L_{\infty }$-algebra and its role in closed string field theory without using operads. 

\vspace{1mm}

{\parindent=0pt 
\underline{Deffinition. (un-shuffle)} }

By a $(k,l)$-unshuffle of $A_{1}, \dots ,A_{n}$ with $n=k+l$ is meant a permutation $\sigma $ such that for $i<j\leq k$, we have $\sigma (i)<\sigma (j)$ and similarly for $k<i<j\leq k+l$. 
We denote the subgroup of $(k,l)$-unshuffle in ${\bf S}_{k+l}$ by ${\bf S}_{k,l}$ and by ${\bf S}_{k+l=n}$, the union of the subgroup ${\bf S}_{k,l}$ with $k+l=n$. 
Similary, a $(k_{1}, \dots , k_{i})$-unsuffle means a permutation $\sigma \in {\bf S}_{n}$ with $n=k_{1}+ \dots + k_{i}$ such that the order is preserved within each block of length $k_{1},\dots ,k_{i}$. 
The subgroup of ${\bf S}_{n}$ consisting of all such unshuffle we denote by ${\bf S}_{k_{1},\dots ,k_{i}}$. 

{\parindent=0pt
\underline{The Koszul Sign of Permutation} }

By decomposing permutations as a product of transpositions, there is then defined the sign of a permutation of $n$ graded elements: $A_{1},\dots ,A_{n}$, e.g for any $\sigma \in {\bf S}_{n}$, the permutation of $n$ graded elements, is defined by 
\begin{align}
\sigma (A_{1} , \dots , A_{n}) = (-1)^{\sigma } ( A_{\sigma (1)}\sigma , \dots , A_{\sigma (n)} ) .
\end{align}
Since we will have many fomulas with such indices and their permutations, we will use the notion $I:=(i_{1}, \dots , i_{n})$ and $A_{I}:=A_{i_{1}} \otimes  \dots \otimes A_{i_{n}}$. Then, for any $\sigma \in {\bf S}_{n}$, we use $\sigma (I)$ to denote $(\sigma (i_{1}), \dots , \sigma (i_{n}))$ and hence $A_{\sigma (I)}=A_{\sigma (i_{1})}\otimes \dots \otimes A_{\sigma (i_{n})}$.

\vspace{1mm}

{\parindent=0pt
\underline{Deffinition. ($L_{\infty }$-algebra)} }

Let $\mathcal{H}$ be a graded vector space and suppose that a collection of degree $(3-2k)$ graded symmetric linear maps ${\bf L}:=\{ L_{k}:\mathcal{H}^{\otimes k}\rightarrow \mathcal{H} \} _{k \geq 0}$ is given. The pair $(\mathcal{H} , {\bf L} )$ is called $L_{\infty }$-algebra if the maps satisfy the following relations: 
\begin{align}
\sum_{\sigma \in S_{k+l=n}} (-1)^{\sigma } L_{l+1} ( L_{k}( A_{\sigma (1)} , \dots , A_{\sigma (k)} ) , A_{\sigma (k+1)} , \dots , A_{\sigma (n)} ) =0 
\end{align}
for $n\geq 1$. A weak $L_{\infty }$-algebra consists of a collection of degree $(3-2k)$ graded symmetric linear maps ${\bf m}:=\{ m_{k}:\mathcal{H}^{\otimes k} \rightarrow \mathcal{H} \} _{k\geq 0}$ satisfying the same relation but for $n\geq 0$ and with $k,l\geq 1$.

{\parindent=0pt
\underline{Deffinition. ($L_{\infty }$-morphism)} }

For two $L_{\infty }$-algebras $(\mathcal{H},{\bf L})$ and $(\mathcal{H}^{\prime } ,{\bf L}^{\prime })$, suppose that there exists a collection of degree preserving graded symmetric multi-linear maps ${\bf F}:= \{ f_{k}:\mathcal{H}^{\otimes k}\rightarrow \mathcal{H}^{\prime } \} _{k\geq 0}$ where $f_{0}$ is a map from $\mathrm{C}$ to a degree zere subvector space of $\mathcal{H}$. 
${\bf F}$ is called an $L_{\infty }$-morphism if it satisfies the following relations
\begin{align}
\sum_{\sigma \in {\bf S}_{k+l=n}} (-1)^{\sigma } f_{1+l} ( L_{k} \otimes {\bf 1}^{\otimes l} ) (A_{\sigma (I)})
=\sum_{\sigma \in {\bf S}_{ k_{1}+ \dots +k_{j}
=n} } \frac{ (-1)^{\sigma } }{ j! } L^{\prime }_{j} ( f_{k_{1}} \otimes \dots  \otimes f_{k_{j}} ) (A_{\sigma (I)})
\end{align}
When $(\mathcal{H} , {\bf L})$ and $(\mathcal{H}^{\prime }, {\bf L}^{\prime })$ are weak $L_{\infty }$-algebra, then a weak $L_{\infty }$-morphism consists of multi-linear maps $\{ f_{k} \} _{k\geq 0}$ satisfying the above condition and $f_{1}\circ L_{0}=\sum_{k} \frac{1}{k!}L^{\prime }_{k}( f_{0},\dots ,f_{0} )$.

\vspace{2mm}

{\parindent=0pt
\underline{The Maurer-Cartan Equation} }

In an $L_{\infty }$-algebra $(\mathcal{H}, {\bf L})$, the Maurer-Cartan equation for $A\in \mathcal{H}$ is given by
\begin{align}
\mathcal{F}(A):=\sum_{k\geq 1} \frac{1}{k!} L_{k} ( A ,\dots , A )=0 .
\end{align}
We donote the set of the solutions of the Maurer-Cartan equation as $\mathcal{MC}(\mathcal{H},{\bf L})$.

Given two elements $\mathcal{G}_{0},\mathcal{G}_{1} \in \mathcal{MC}(\mathcal{H},{\bf L})$ are called gauge equivalent if there exists a piecewise smooth path $\mathcal{G}(\tau )$, $\tau \in [0,1]$ such that 
\begin{align}
\frac{d}{d\tau }\mathcal{G}(\tau ) = \sum_{k\geq 0} \frac{1}{k!} L_{k+1} ( \Lambda (\tau ), \mathcal{G}(\tau ) , \dots , \mathcal{G}(\tau )  )
\end{align}
for a degree $\mathrm{deg}[\mathcal{G}(\tau )]-1$ element $\Lambda (\tau )$ where $\mathcal{G}(0)=\mathcal{G}_{0}$ and $\mathcal{G}(1)=\mathcal{G}_{1}$. 
Gauge transformation preserves $\mathcal{MC}(\mathcal{H},{\bf L})$ and defines the equivalence relation. 
Then a quotient space by this equivalence relation: $\mathcal{M}(\mathcal{H},{\bf L}) := \mathcal{MC}(\mathcal{H},{\bf L})/{\sim  }$ gives the moduli space of the deformations.

{\parindent=0pt
\underline{Null forms of the Maurer-Cartan operator} }

For $A\in \mathcal{MC}(\mathcal{H},{\bf L})$ and $B\in \mathcal{H}$ satisfying $\mathrm{deg}[B]=\mathrm{deg}[A]-1$, there exist special element:
\begin{align}
\mathcal{L}_{A}(B):=\sum_{k\geq 0} \frac{1}{k!} L_{k+1} ( A,\dots , A ,B ) .
\end{align}
It gives a null state in the Maurer-Cartan operator, or conversely, the image of the Maurer-Cartan perator is included in the kernel of this operator: $\mathcal{L}_{A}(\mathcal{F}(A))=0$. 

\vspace{2mm}

{\parindent=0pt
\underline{Symplectic Structure} }

Skew-symmetric bilinear map $\omega : \mathcal{H}\otimes \mathcal{H}\rightarrow \mathrm{C}$ is called constant symplectic structure when it has fixed integer degree $|\omega |\in \mathrm{Z}$ and is non-degenerate. Namely, 
\begin{align}
\omega (A, B) =-(-1)^{AB} \omega (B,A)
\end{align}
for any $A,B \in \mathcal{H}$ and degree $|\omega |$ implies that $\omega (A,B)=0$ except for $|A|+|B|+|\omega |=0$. 

{\parindent=0pt 
\underline{Cyclic $L_{\infty }$-algebras} }

Suppose that an $L_{\infty }$-algebra $(\mathcal{H},{\bf L})$ is equipped with constant symplectic structure $\omega :\mathcal{H}\otimes \mathcal{H}\rightarrow \mathrm{C}$. 
For ${\bf L}= \{ L_{k}:\mathcal{H}^{\otimes k}\rightarrow \mathcal{H} \}$, let us define the multi-linear maps by
\begin{align}
\mathcal{V}_{n+1} ( A_{0},A_{1},\dots A_{n} ) :=\omega ( A_{0} , L_{n}( A_{1}, \dots , A_{n} ) ) .
\end{align}
The degree of $\mathcal{V}_{n+1}$ is $|\omega |+1$. 
An $L_{\infty }$-algebra equipped with constant symplectic structure $(\mathcal{H}, {\bf L}, \omega )$ is called cyclic $L_{\infty }$-algebra if $\mathcal{V}_{n+1}$ is graded symmetric with respect to any permutation of $\mathcal{H}^{\otimes (n+1)}$:
\begin{align}
\mathcal{V}_{n+1} ( A_{0}, \dots , A_{n} ) =(-1)^{\sigma } \mathcal{V}_{n+1} ( A_{\sigma (0)}, \dots , A_{\sigma (n)} )  , \hspace{5mm} \sigma \in S_{n+1} . 
\end{align}
In string field theory, we sometimes use the Stokes's Teorem: 
\begin{align}
\sum_{\mathcal{V}} \sum_{\sigma \in {\bf S}_{n+1}} (-1)^{\sigma } \mathcal{V}_{n+1}( L_{1}(A_{\sigma (0)}), A_{\sigma (1)}, \dots ,(A_{\sigma (n)}) ) =0 .
\end{align}
If the Stokes's Theorem holds in cyclic $L_{\infty }$-algebra, then $L_{\infty }$-identities splits into two groups: 
\begin{align}
L_{1}(L_{n}(A_{1}, \dots , A_{n}) ) +
\sum_{\sigma \in {\bf S}_{n}} (-1)^{\sigma } L_{n}( L_{1}(A_{\sigma (1)}), A_{\sigma (2)}, \dots ,(A_{\sigma (n)}) ) &=0 ,
\\
\sum_{\sigma \in {\bf S}_{k+l=n}} (-1)^{\sigma } L_{l+1} ( L_{k}( A_{\sigma (1)} , \dots , A_{\sigma (k)} ) , A_{\sigma (k+1)} , \dots , A_{\sigma (n)} ) =& \, 0 ,
\end{align}
where $l\geq 1$ and $k\geq 2$. Then $L_{1}$ becomes the derivation for all products $\{ L_{k} \} _{k\geq 2}$. 
It is naturally relized in considering scattering amplitudes of multi-strings in closed string field theory. 

\vspace{2mm}

{\parindent=0pt{ \underline{$L_{\infty }$-algebras and Cyclicity in Closed Sting Field Theory} } }

\vspace{1mm}

In string field theory, these mathematical objects naturally corespond to physical ones. 
We often use $L_{\infty }$-identities by the following form: ($l\geq 1$, $k\geq 2$ and $\sigma $ is the sign of splittings) 
\begin{align}
\sum_{l+k=n} \sigma (i_{l},i_{k}) [A_{i_{1}}, \dots ,A_{i_{l}} , [ A_{j_{1}} , \dots , A_{j_{k}} ] ]=0 .
\end{align}
The sum runs over all different splittings of the set $\{ 1, \dots , n \}$ into a first group $\{ i_{1}, \dots ,i_{l} \}$ and a second group $\{ j_{1} , \dots , j_{k} \}$, where $l\geq 1$ and $k \geq  2$. 
Two splittings are the same if the corresponding first groups contain the same set of integers regardless of their order. 
The sign factor $\sigma (i_{l}, j_{k})$ 
is defined to be the sign picked up when one rearranges the sequence $\{ Q, A_{1}, A_{2}, \dots , A_{n} \}$ into the sequence $\{ A_{i_{1}}, \dots , A_{i_{l}}, Q, A_{j_{1}}, \dots , A_{j_{k}} \}$ taking into account the Grassmann property. 
The gauge structure of closed string theory is governed by $L_{\infty }$-algebras. 
In particular, the Maurer-Cartan equation corresponds to the equation of motion $\mathcal{F}(\Psi )=0$ and the null form $\mathcal{L}_{\Psi }(\Lambda )$ corresponds to the gauge transformation $\delta \Psi =Q_{\Psi }\Lambda $. 
The differential equation which defines a pure gauge solution comes from above equation which decides the gauge equivalence of Maurer-Cartan elements. 
In addition to these, the symplectic form $\omega (A,B)$ is given by the BPZ inner product $\langle A | c^{-}_{0} | B \rangle $ of conformal field theory. 
The cyclic $L_{\infty }$-structure $\{ \mathcal{V}_{n} \}_{n}$ corresponds to the set of fundamental vertices of string field theory. 
Furthermore, in string field theory, there exists Stokes's Theorem on the moduli space of Riemann surfaces. 
Therefore, closed string field theory is described by cyclic $L_{\infty }$-algebras equipped with Stokes's Theorem. 

\vspace{1mm}

{\parindent=0pt
\underline{Stokes's Theorem on Moduli Space in Closed String Field Theory} }

\vspace{1mm}

The closed string products are constructed form the decomposition of the moduli space of Riemann surfaces\cite{Zwiebach:1992ie}. 
We know the fact that the BRST operator acts on the moduli sapce $\mathcal{M}_{n}$ as an exterior product acting on the cotangent bundle of $(2n-6)$-dimensional manifold \cite{Kimura:1993ea} and there also exists good theorem, so-called Stoke's Theorem. 
The $n$-point correlation function $\Omega _{A_{1} \dots A_{n}}$ of string field theory corresponds to the top form $\Omega ^{[0]}$ on the moduli space $\mathcal{M}_{n}$. 
As a result, we obtain 
\begin{align}
\int_{\mathcal{M}_{n+1}}\Omega _{(\sum Q)A_{0} \dots A_{n}}^{[0]}
= \int_{\mathcal{M}_{n+1}} d \, \Omega _{A_{0} \dots A_{n}}^{[-1]}
= \int_{\partial \mathcal{M}_{n+1}} \Omega _{A_{0} \dots A_{n}}^{[-1]}=0
\end{align}
where $\mathcal{M}_{n+1}$ is the moduli space of the $(n+1)$-punctured sphere and $\Omega ^{[-r]}$ is the volume form of such a space whose dimension is $(\mathrm{dim}[\mathcal{M}_{n+1}]-r)$. 
By definition, $(n+1)$-point correlation functions $\mathrm{A}( A_{0}, \dots ,A_{n} ) $ are given through the integral of the top form over such a moduli space $\mathcal{M}_{n+1}$. 
\begin{align}
\sum_{\mathrm{graph}}\sum_{i=0}^{n} (-1)^{A_{0}+\dots +A_{i-1}} \langle A_{0} , [ A_{1} , \dots ,QA_{i} , \dots A_{n} ] \rangle &=\int_{\mathcal{M}_{n+1}}\Omega _{(\sum Q)A_{0} \dots A_{n}}^{[0]}
\end{align}
The summention runs over decomposition of the moduli space $\mathcal{M}_{n}\equiv \mathcal{V}_{n}\oplus R_{n}^{1}\oplus \dots \oplus R_{n}^{n-3}$, where $\mathcal{V}_{n}$ gives the $n$-point fundamental vertex and $R_{n}^{i}$ represents the region of $n$-point graph vertices constructed from the possible combination of lower vertecies and propergators. 
Therefore, 
\begin{align}
\sum_{\mathrm{graph}} \Big( 
\langle Q A_{0} , [ A_{1}, \dots , A_{n} ] \rangle 
+  \sum_{i=1}^{n} (-1)^{A_{0}+A_{1}+ \dots + A_{i-1}} \langle  A_{0} , [ A_{1} , \dots , QA_{i} , \dots , A_{n} ] \rangle  
\Big) =0.
\end{align}
Note that in the case of the three point vertex, amplitudes are obtained by the fundamental vertex only, namely, $\mathcal{M}_{3}\equiv \mathcal{V}_{3}$. Then BRST-exact terms decouple.

\section{A Parallel Structure of Actions} \label{Appendix B}

{\parindent=0pt
\underline{Free Theory of Bosonic Closed Strings} }

\vspace{1mm}

We already know the free action for bosonic closed string field theory: $S_{0}=\frac{1}{2}\langle \Psi , Q \Psi \rangle$. 
Then the generator of the gauge transformation is given by $Q$. 
The state $Q^{\prime }\Psi $ which belongs to the kernel of this generator $Q$ is given by $Q^{\prime }\Psi := Q\Psi $ because of the nilpotency $Q^{2}=0$. 
Therefore, in free theory, we can regard the Maurer-Cartan operator $Q^{\prime }$ as the BRST operator $Q$ itself. 
Using this $Q^{\prime }$, we can rewrite the free action of bosonic string field theory as follows,
\begin{align}
S_{0}=\int_{0}^{1} {dt} \, \langle \partial _{t} \Psi (t) , Q^{\prime }\Psi (t) \rangle ,
\end{align}
where $\Psi (t)$ satisfies $\Psi (0)=0$ and $\Psi (1)= \Psi $ for $t \in [0,1]$ and $Q^{\prime }\Psi \equiv Q \Psi $. 
For a given derivation $X$ satisfying $[[Q,X]]=0$, the state $Q^{\prime }\Psi $ satisfies $XQ^{\prime }\Psi =XQ\Psi = (-1)^{X}QX \Psi $. 
Thus we can easily calculate the variation as follows:
\begin{align}
\nonumber 
\delta S_{0}&= \int_{0}^{1}{dt} \, \Big( \langle \partial \delta \Psi (t), Q^{\prime }\Psi (t) \rangle + \langle \partial \Psi (t) , Q \delta \Psi (t) \rangle  \Big) 
\\ \nonumber 
&= \int_{0}^{1}{dt}  \, \Big  \langle \partial _{t} \delta \Psi (t), Q^{\prime } \Psi (t) \rangle + \langle Q \partial _{t} \Psi (t) , \delta \Psi (t) \rangle \Big)
\\
&=\int_{0}^{1}{dt} \, \partial _{t} \, \langle \delta \Psi (t) , Q^{\prime } \Psi (t) \rangle = \langle \delta \Psi , Q^{\prime } \Psi \rangle .
\end{align}
In full theory, the generator of the gauge transformation $Q_{\Psi }$ does not have the nilpotency: we obtain $Q_{\Psi }^{2}=0$ if and only if $\Psi$ satisfies the equation of motion $Q^{\prime }_{\Psi }\Psi=0$. 
For a general string field $\Psi$, the state which belongs to the kernel of $Q_{\Psi}$ is given by $Q^{\prime }_{\Psi }\Psi $ from $L_{\infty }$-identities: $Q_{\Psi }Q_{\Psi }^{\prime }\Psi =0$. 
Thus $Q^{\prime }_{\Psi }\not= Q_{\Psi }$ in full theory. 

\vspace{2mm}

{\parindent=0pt
\underline{The Difference for Bosonic and Type II Theories} }

\vspace{1mm}

In contrast to calculations (\ref{bosonic calculation}) in bosonic theory, $[\delta \Psi  , \partial _{t} \Psi  ]^{[a]}_{\Psi }$ and $[ \partial _{t} \Psi   , \delta \Psi  ]^{[a]}_{\Psi }$ in (\ref{bosonic calculation}) are replaced by $[\delta \Psi , \partial _{t} \mathcal{G} ]_{\mathcal{G}}$ and $[\partial _{t} \Psi , \delta \mathcal{G}]_{\mathcal{G}}$ in type II theory. 
However, these terms do not cancel each other but vanish respectively. 
If the following relation holds, there would exist cancellation 
\begin{align}
[\partial _{t}\Psi (t) , \delta \mathcal{G}(t) ]_{\mathcal{G}} \overset{?}{=} [ \delta \Psi (t) , \partial _{t} \mathcal{G} (t) ]_{\mathcal{G}} .
\end{align}
This relation is trivial for the case like analytic functions, graded differential algebras, and so on because of the associativity of the product. 
A typical example is a analytic function $f[z(t)]$:
\begin{align}
\delta f[z(t)] \cdot \partial _{t}z(t) = \partial _{t} f[z(t)]\cdot \delta z(t) .
\end{align}
Unfortunately, this relation does not hold for a general element of $L_{\infty }$-algebras. 
However, considering the property of the BPZ inner product, we notice that such extra elements belong to the kernel of cyclic $L_{\infty }$-algebras. 
The decoupling mechanism of BRST-exact states works. 

\vspace{1mm}

There is a slight difference between bosonic action and our type II action and it is a result from that we use a naive extension of Zwiebach's string products which is constructed from the decomposition of bosonic moduli space and carry appropriate ghost numbers for bosonic string fields. 
We expect that after constructing type II vertices as the decomposition of the muduli space and identifying some underlying (homotopy) algebra, using such vertices for type II theory, we will be able to write down the action which have exactly parallel structure.






\section{Open String Field Theory} \label{Appendix C}
Open string field theory \cite{Witten:1985cc} is described by the Chern-Simons-like action
\begin{align}
S_{cs}=-\frac{1}{g^{2}} \left( \frac{1}{2}\langle \Phi ,Q\Phi \rangle +\frac{g}{3}\langle \Phi , \Phi \ast \Phi \rangle \right) .
\end{align}
Thus the equation of motion is given by $Q\Phi + \frac{1}{2}\Phi \ast \Phi =0$ and the gauge invariance is generated by $Q+[[g\Phi , \hspace{2mm} ]]$. 
Here, $[[A,B]]$ means the graded commutater: $[[A,B]]:=A\ast B-(-1)^{AB}B\ast A$. 
In this theory, the gauge structure is governed by differential graded algebras, which are the special case of $A_{\infty}$ algebras \cite{Gaberdiel:1997ia, Nakatsu:2001da, Kajiura:2001ng}. 
We can also rewrite this action into our form: $S=\int{dt}\langle \partial _{t}\Phi ,Q_{\Psi }\Phi \rangle$.

Let us define the BRST operator around $A$ with coupling constant $ag$ and the Maurer-Cartan operator $Q^{\prime }_{A}$ as follws: 
\begin{align}
Q_{A}^{[a]} \Phi  :=Q\Phi + ag[[ A , \Phi ]] ,
\hspace{5mm}
Q^{\prime }_{A} \Phi := \int_{0}^{1}{da} \, Q_{A }^{[a]}\Phi = Q\Phi + \frac{g}{2} [[ A, \Phi ]] . 
\end{align}
Then we can rewrite this Charn-Simons-like action into the following form:
\begin{align}
\nonumber 
S_{cs}=& -\frac{1}{g^{2}} \left( \frac{1}{2}\langle \Phi ,Q\Phi \rangle +\frac{g}{3!}\langle \Phi ,[[ \Phi , \Phi ]] \rangle \right) 
\\ \nonumber 
=& -\frac{1}{g^{2}} \int_{0}^{1}{dt} \langle \partial _{t} (t \Phi ) ,Q(t\Phi )+ \frac{g}{2}[[ (t\Phi ), (t\Phi ) ]] \rangle
\\
=& -\frac{1}{g^{2}} \int_{0}^{1}{dt} \langle \partial _{t} \Phi (t) , Q^{\prime }_{\Phi (t)} \Phi (t)  \rangle ,
\end{align}
where $t\in [0,1]$ is a real parameter. 
At the last line, we introduce $\Phi (t)$ which satisfies $\Phi (0)=0$ and $\Phi (1)=\Phi $. 
Then we notice that the equation of motion is give by the Maurer-Cartan operator: $Q^{\prime }_{\Phi }\Phi =0$ and the gauge invariance is generated by $Q_{\Phi }$. 
(Of course, when $A$ satisfies the equation of motion in open string field theory, $Q_{A}$ becomes nilpotent operator.)

\vspace{2mm}

{\parindent=0pt{\underline{Open Superstring Field Theory} }}

\vspace{1mm}

Recall that using a gauge parameter field of bosonic theory $\lambda $ whose ghost number is $0$, we can construct a pure gauge solution of boconic theory by $A_{Q}:=e^{-\lambda }(Qe^{\lambda })$. 
If there exist a derivation $X$ which satisfies $[[Q,X]]=0$, a related field $A_{X}:=e^{-\lambda }(Xe^{\lambda })$ which satisfies $ (-1)^{X} X A_{Q}=Q_{A_{Q}} A_{X}\equiv QA_{X}+g[[A_{Q} , A_{X} ]]$ appears. 
In the large Hilbert space description, replacing these parameter fields $\lambda$ by superstring fields $\Phi (t)$ which satisfy $\Phi (0)=0$ and $\Phi (1)=\Phi $, we obtain the following action of Berkovits' theory \cite{Berkovits:1995ab}:
\begin{align}
S_{wzw}=\int_{0}^{1}{dt} \langle \eta _{0} A_{\partial _{t}}(t), A_{Q}(t) \rangle \overset{\Phi (t) =t\Phi }{=}\int_{0}^{1}{dt} \langle \eta_{0} \partial _{t}(t \Phi ) ,A_{Q}(t) \rangle . \label{wzw}
\end{align}
The equation of motion is given by $\eta _{0}A_{Q}=0$ and this action is invariant under $A_{\delta }=Q_{A_{Q}}\Lambda $. 
We would like to mention that for linear $t$-dependent $\Phi (t)$, introducing a real parameter $a\in [0,1]$, we can rewrite this action into our new form: 
\begin{align}
S_{wzw}= \int_{0}^{1}{dt} \langle  \eta _{0} \partial _{t}\Phi (t), Q ^{\prime }_{A_{Q}(t)} \Phi (t) \rangle , \label{another}
\end{align}
where $Q^{\prime }_{A_{Q}}$ is defined by the $a$-integration of $Q_{A_{Q}}^{[a]}$ from $0$ to $1$. 
Note that now $Q^{[a]}_{A_{Q}}$ is given by 
\begin{align}
Q_{A_{Q}(t)}^{[a]} = Q \phi + [[ A^{[a]}_{Q}(t) , \hspace{2mm} ]] ,
\hspace{4mm}
A_{Q}^{[a]}(t) \equiv  e^{-ag \Phi (t)} \left( Qe^{ag \Phi (t)} \right) .
\end{align}
The equivalence of (\ref{wzw}) and (\ref{another}) is provided by the relation $A_{Q}(t)=Q_{A_{Q}(t)}^{\prime }\Phi (t)$ for $\Phi (t) = t\Phi$. 
Let us check this relation. 
Using $\Phi = g\tilde{\Phi }$, we can rewrite $A_{Q}$ into $Q^{\prime }_{A_{Q}}\Phi$ as follows:
\begin{align}
\nonumber 
A_{Q}(t)=&e^{-t\Phi }\left( Q e^{t\Phi }  \right) 
\\ \nonumber 
=& \, tQ\Phi  + \frac{t^{2}}{2}(Q\Phi \ast \Phi -\Phi \ast Q\Phi  ) 
+\frac{t^{3}}{3!}( Q\Phi \ast \Phi ^{2} -2\Phi \ast Q\Phi \ast \Phi + \Phi ^{2}\ast Q\Phi  ) 
\\ \nonumber 
& \,\, +\frac{t^{4}}{4!}(Q\Phi \ast \Phi ^{3} -3\Phi \ast Q\Phi \ast \Phi ^{2} +3\Phi ^{2}Q\Phi \ast \Phi -\Phi ^{3}Q\Phi  ) 
+O(t^{5}) 
\\ \nonumber 
=& \sum_{n=0}^{\infty } \frac{t^{n}}{n!}\sum_{i+j=n-1} \frac{(n-1)!}{i!j!}\Phi ^{i} \ast Q\Phi \ast \Phi ^{j}
= \sum_{n=0}^{\infty }\frac{t^{n}}{n!} \big[ \big[ \dots [[ Q \Phi ,\Phi ]] , \Phi ]], \dots , \Phi \big] \big] 
\\ \nonumber 
=& \, Q(t\Phi ) + \int_{0}^{1}{da} \sum_{n=2}^{\infty }\frac{(ag)^{n-1}t^{n}}{(n-1)!} \big[ \big[ \dots [[ Q \tilde{\Phi },\tilde{\Phi } ]] , \tilde{\Phi }]], \dots , \tilde{\Phi }] ] ,\Phi \big] \big] 
\\
=& \, Q\Phi (t) + \int_{0}^{1}{da} \, [[ A_{Q}^{[a]}, \Phi (t) ]]=Q_{A_{Q}(t)}^{\prime } \Phi (t) .
\end{align}

\parindent=0pt

\end{document}